\documentclass[numbook,envcountsame]{svjour2}
 \smartqed
 \usepackage{graphicx,amsfonts,amsmath,amssymb,latexsym,cancel}
 \journalname{Journal of Statistical Physics}

 \newcommand\sn{\mathrm{sn}}
  \newcommand\cn{\mathrm{cn}}
   \newcommand\dn{\mathrm{dn}}

 \newcommand\Rb{\mathbb{R}}
 \newcommand\Cb{\mathbb{C}}

 \newcommand\psid{\psi^{\dag}}

 \newcommand\kx{\mathcal{K}_x}
 \newcommand\ky{\mathcal{K}_y}

 \newcommand\ben{\begin{equation*}}
 \newcommand\ebn{\end{equation*}}
 \newcommand\be{\begin{equation}}
 \newcommand\eb{\end{equation}}

  \newcommand\smsum{\text{\footnotesize $\sum$}}
 \newcommand{\pf}{\begin{bpf}}

 \newcommand{\pfms}{\begin{bpfms}}
 \newcommand{\epf}{\end{bpf}\hfill$\square$\vspace{0.1cm}}
 \newcommand{\epfms}{\end{bpfms}\hfill$\square$\\ }

 \title{Ising correlations and elliptic determinants}
 \begin{document}
 \author{N. Iorgov \and O. Lisovyy}
 \institute{N. Iorgov \at Bogolyubov Institute for Theoretical Physics, 03680 Kyiv, Ukraine
   \\ Max-Planck-Institut f\"ur Mathematik, 53111 Bonn, Germany \\
 \email{iorgov@bitp.kiev.ua}
 \and O. Lisovyy \at Laboratoire de Math\'ematiques et Physique Th\'eorique CNRS/UMR 6083, \\
  Universit\'e de Tours, Parc de Grandmont, 37200 Tours, France \\
 \email{lisovyi@lmpt.univ-tours.fr} }
 \date{Received: date / Accepted: date}

 \maketitle
 \begin{abstract} Correlation functions of the two-dimensional Ising model on the periodic
 lattice can be expressed in terms of form factors --- matrix elements of the spin operator in the basis
 of common eigenstates of the transfer matrix and translation operator. Free-fermion
 structure of the model implies that any multiparticle form factor is given by the pfaffian
 of a matrix constructed from the two-particle ones. Crossed two-particle form factors
 can be obtained by inverting a block of the matrix of linear transformation induced on fermions by the spin conjugation.
 We show that the corresponding matrix is of elliptic Cauchy type and use this observation
 to solve the inversion problem explicitly. Non-crossed two-particle form factors are then obtained
 using theta functional interpolation formulas. This gives a new simple proof of the factorized
 formulas for periodic Ising form factors, conjectured by A.~Bugrij and one of the authors.
 \end{abstract}

 \section{Introduction}
 The two-dimensional Ising model is the foundation of our understanding of critical phenomena and quantum
 field theory. Starting from Onsager's calculation of the partition function \cite{Onsager}, its
 study has led to remarkable developments in many areas of mathematical physics, including
 the theory of Toeplitz matrices, quantum integrable systems, Painlev\'e equations, and
 conformal field theory. In an effort to keep the bibliography of reasonable size, we refer the reader to \cite{McCoy,McCoyWu,Palmer_book} for an account of some of these developments and further references.

 Correlation functions of the Ising model on an $M\times N$ square lattice with periodic or antiperiodic
 boundary conditions in each direction can be expressed in terms of elementary blocks,
 called form factors. These blocks are matrix elements of the spin operator in the basis
 of common eigenstates of the $2^N\times 2^N$ transfer matrix and the operator of translations.
 Compact factorized formulas for the form factors of Ising spin have been conjectured in \cite{BL1,BL2}
 on the basis of (i) long-distance expansions of the two-point correlation function on the
 cylinder ($M\rightarrow\infty$) \cite{Bugrij} and (ii) direct transfer matrix calculations for small values of
 $N$.

 One can try to prove the conjectures of \cite{BL1,BL2} in several ways, reflecting
 different facets of the integrable structure of the Ising model. First, it can be seen as
 a special case of the eight-vertex model. This is not very helpful, as the methods
 of computation of finite-lattice form factors for models with elliptic $R$-matrix have
 not yet been developed.
 Another option is to consider the Ising model as a particular case
 of the Baxter-Bazhanov-Stroganov $\tau^{(2)}$-model \cite{Baxter,BS,Korepanov}. The corresponding
 $R$-matrix is trigonometric, but the $L$-operator intertwines spin-$\frac12$ and
 cyclic type evaluation representations
 of the quantum affine algebra $U_q(\widehat{sl_2})$. For this reason, the transfer matrix
 can not be diagonalized by the standard algebraic Bethe ansatz technique,
 the corresponding quantum inverse scattering problem does not
 admit a simple solution, and the Lyon method of computation of form factors for XXZ-type
 models \cite{xxz} can not be applied.

 These difficulties can be partially overcome in the framework of Sklyanin's method of separation
 of variables \cite{Sklyanin}. Local spin operators are then expressed
 in terms  of the quantum monodromy matrix by means of an iterative procedure.
 The transfer matrix eigenstates are represented
 by linear combinations of eigenvectors of an auxiliary problem,
 and the computation of form factors involves summations over this auxiliary basis. Even in the Ising case, performing
 the sums explicitly and showing that they do reduce to factorized expressions
 of \cite{BL1,BL2} is a highly nontrivial problem, which was solved in \cite{Iorgov1,Iorgov2}. In spite of
 this success, the whole procedure looks overwhelmingly cumbersome as compared to
 final answer, and there remains a strong feeling that
 the result can (and has to) be obtained in a much simpler way.

 The most natural approach would be to take advantage of the free-fermion structure of the Ising
 model. This structure was discovered in \cite{kaufman}, where it was used to give an alternative derivation of the partition function.

 Recall that the periodic Ising transfer matrix $V$ commutes
 with the global spin flip operator $U$. Since $U^2=\mathbf{1}$, the eigenstates of
 $V$ split into two sets, corresponding to $+1$ and $-1$ eigenvalues of $U$.
 The transfer matrix can be naturally written as
 \be\label{vap}
 V=\mathrm{const}\cdot\left(\frac{\mathbf{1}+U}{2}\,V_a+\frac{\mathbf{1}-U}{2}V_p\right),
 \eb
 where $V_{a,p}$  commute with the projectors $\displaystyle\frac{1\pm U}{2}$
 and are expressed in terms of the generators of a Clifford algebra. They actually belong
 to the Clifford group, i.e. the generators transform linearly under conjugation by $V_{a,p}$. We
 emphasize that the same is \textit{not} true for the full transfer matrix $V$.

 Clifford group elements can, in principle, be reconstructed from the induced linear transformations, see e.g. Chapter~II in \cite{Berezin}. The case of $V_{a,p}$ is especially simple, as these matrices act
 on the generators as  products of commuting two-dimensional
 rotations; this becomes manifest after
 discrete Fourier transforms (antiperiodic for $V_a$ and periodic for $V_p$). Diagonalization
  of these rotations gives the spectrum of $V_{a,p}$ and hence,
  by (\ref{vap}), that of $V$. The eigenvectors of $V_a$ and $V_p$ are given by multiparticle fermionic Fock states, constructed using two different sets of the creation-annihilation operators \cite{kaufman}.

  Ising spin operator $\sigma$ also belongs to the Clifford group.
  However, it changes the value of $\mathbb{Z}_2$-charge $U$ from $+1$ to $-1$
  and vice versa. According to (\ref{vap}), these values are associated
  with fermionic eigenstates of $V$ of  different types, and it is not clear how one
  can compute form factors between them.
  The problem can be circumvented on the infinite lattice ($N\rightarrow\infty$), where
  the distinction between two types of fermions effectively disappears.
  In that case, two-particle Ising spin form factors have been
  found in \cite{smj}, Theorems~5.2.1--5.2.3, and multiparticle ones in
  \cite{PT}, Theorem~5.0.

  Further progress for finite $N$ was made by Hystad and Palmer
  in two recent papers \cite{Hystad,Palmer_Hystad}.
  It was noticed that if we combine the
  creation-annihi\-la\-tion operators of $a$- and $p$-fermions into  $N$-dimensional
  column vectors
  $\vec{\psi}^{\,\dag}$, $\vec{\psi}$, ${\vec{\varphi}}^{\,\dag}$, $\vec{\varphi}$, and express the result of conjugation of $p$-fermions by $\sigma$ in the $a$-basis,
  \ben
  \sigma\left(\begin{array}{l}\vec{\varphi}^{\,\dag} \\ \vec{\varphi} \end{array}\right)
  \sigma^{-1} =
  \left(\begin{array}{cc} A & B \\ C & D\end{array}\right)
  \left(\begin{array}{l}\vec{\psi}^{\,\dag} \\ \vec{\psi} \end{array}\right),
  \ebn
  then form factors of $\sigma$ between Fock states of different types
  can be expressed in terms of induced rotations in essentially the same way
  as if we had
  an ordinary Bogoliubov transformation involving fermions of only one type. For example, for $T<T_c$
  \begin{itemize}
  \item an arbitrary $n$-particle form factor is
  given by the pfaffian of an $n\times n$ matrix, constructed from the
  two-particle ones;
  \item normalized two-particle form factors coincide with the elements
  of $N\times N$ matrices $D^{-1}$, $BD^{-1}$, $D^{-1}C$.
  \end{itemize}
  Explicit form of the blocks $A$, $B$, $C$, $D$ can be fixed in a rather
  straightforward way. The computation of finite-lattice form factors
  therefore reduces to the inversion of $D$ and calculation
  of the matrix products $BD^{-1}$ and $D^{-1}C$. The present
  work is devoted to the solution of these two problems.

  The key point of our analysis is an elliptic pa\-ra\-metrization of the Ising
  spectral curve, equivalent to the elliptic substitutions used in Yang's derivation
  of the spontaneous magnetization \cite{Yang}. We will show that in this pa\-ra\-me\-trization
  $D$ is given by an elliptic  Cauchy matrix. Its determinant and inverse can be computed
  using the so-called Frobenius determinant formula \cite{Frobenius}.
  (Let us mention that
  the Frobenius determinant and its generalizations have recently appeared in the computation of the partition function
  of the eight-vertex SOS model with domain wall boundary conditions \cite{Rubtsov,Rosengren1}).
  In fact taking the $N\rightarrow\infty$ limit of
  $\mathrm{det}\,D$ provides a useful alternative to Yang's derivation. Further, the products
  $BD^{-1}$, $D^{-1}C$ can be calculated using theta functional analogs of the Lagrange interpolation identities.
  The matrix of two-particle form factors also turns out to be of elliptic Cauchy type. Computing
  its pfaffian and going back to the usual trigonometric parametrization, we obtain
  the general formula for finite-lattice form factors of Ising spin, conjectured in \cite{BL1,BL2}.

  It is worth noticing that in a recent paper \cite{Iorgov3}
  form factors of the quantum Ising chain in a transverse field were derived.
  They represent a limiting case of the form factors obtained in the present work.
  Although the derivation in \cite{Iorgov3} also uses the fermion algebra,
  it is motivated by the calculation of form factors for superintegrable chiral Potts quantum
  chain \cite{Iorgov4} using Baxter's extension \cite{Baxter2} of Onsager algebra
  and it does not use  linear transformations induced by the spin operator.

 This paper is organized as follows. In Section~\ref{sec_notation}, we
 introduce basic notation and recall Kaufman's fermionic approach \cite{kaufman}
 to the diagonalization of the Ising transfer matrix. Section~\ref{sec_rotations} translates
 the results of \cite{Hystad,Palmer_Hystad}, expressing spin form factors in terms of
 induced linear transformations, into the language of creation-annihilation
 operators. In Section~\ref{elpar}, we derive a number of identities relating
 trigonometric and elliptic parametrization of the Ising spectral curve. The main result is Lemma~\ref{lemsff},
 which gives elliptic representations for the
 matrix elements of $A$, $B$, $C$ and $D$. Section~\ref{sec_cauchy}
 is devoted to the computation of two-particle form factors, i.e. the matrices $D^{-1}$, $BD^{-1}$ and $D^{-1}C$.
 The principal results of this section are summarized in Theorem~\ref{thm2pff}. Finally,
 the general factorized formula for the multiparticle form factors is established
 in Section~\ref{gfsection}. We conclude with a brief discussion of results and open questions.

 \section{Transfer matrix diagonalization\label{sec_notation}}
 The two-dimensional Ising model on an $M\times N$ lattice is described by the following Hamiltonian:
 \ben
 -\beta H[\sigma]=\sum_{j=0}^{M-1}\sum_{k=0}^{N-1}\Bigl(\kx \sigma_{j,k}\sigma_{j+1,k}+\ky\sigma_{j,k}\sigma_{j,k+1}\Bigr).
 \ebn
 Spin variables $\{\sigma_{j,k}\}_{\substack{j=0,\ldots,M-1\\k=0,\ldots,N-1}}$ take on the values $\pm1$,
 and the boundary conditions in each direction are  either periodic or antiperiodic. This means that
 for $j=0,\ldots,M-1$ and $k=0,\ldots,N-1$ we have
 $ \sigma_{M,k}=\varepsilon_x \sigma_{0,k}$ and
 $\sigma_{j,N}=\varepsilon_y \sigma_{j,0}$
 with $\varepsilon_{x,y}=\pm1$. To simplify some of the arguments below, it will be assumed
 that $\mathcal{K}_{x,y}\in\mathbb{R}_{\geq0}$. From physical point of view, this involves
 no loss of generality, as any real values of coupling constants can be made non-negative
 by a suitable change of spin variables and boundary conditions.

 Typically one is interested in calculating $n$-point correlation functions
 \be\label{corfun}
 \langle\sigma_{j_1,k_1}\ldots\sigma_{j_n,k_n}\rangle=\frac{\sum_{[\sigma]}
 \sigma_{j_1,k_1}\ldots\sigma_{j_n,k_n}e^{-\beta H[\sigma]}}{\sum_{[\sigma]}e^{-\beta H[\sigma]}}.
 \eb

 The $2^N\times 2^N$ transfer matrix $V$ can be thought of as an operator of discrete evolution
 in the $x$-direction. It acts in the space of states, which admits two equivalent
 descriptions: as a space of maps $f:\left(\mathbb{Z}_2\right)^{\times N}\rightarrow\mathbb{C}$ or as an $N$-fold tensor product
 $\mathbb{C}^2\otimes \ldots\otimes \mathbb{C}^2$. To write $V$ explicitly,
 it is common to define the operators
 \begin{align*}
 s_j=&\,\underbrace{\mathbf{1}\otimes\ldots\otimes\mathbf{1}}_{j\;\text{times}}\otimes
 \,\sigma_z\otimes\mathbf{1}\otimes
 \ldots\otimes\mathbf{1},\\
 C_j=&\,\underbrace{\mathbf{1}\otimes\ldots\otimes\mathbf{1}}_{j\;\text{times}}\otimes
 \,\sigma_x\otimes\mathbf{1}\otimes
 \ldots\otimes\mathbf{1},
 \end{align*}
 where  $j=0,\ldots,N-1$ and $\sigma_{x,y,z}$ denote the Pauli matrices
 \ben
 \sigma_x=\left(\begin{array}{cc} 0 & \;1 \\ 1 & \;0 \end{array}\right),\qquad
 \sigma_y=\left(\begin{array}{cr} 0 &-i \\ i & 0 \end{array}\right),\qquad
 \sigma_z=\left(\begin{array}{cr} 1 & 0 \\ 0 &-1 \end{array}\right).
 \ebn
 The action of $\{s_j\}$, $\{C_j\}$ on the space of maps is given by
 \begin{align*}
 \left(s_jf\right)(\sigma_0,\ldots,\sigma_{N-1})=&\,\sigma_j f(\sigma_0,\ldots,\sigma_{N-1}),\\
 \left(C_jf\right)(\sigma_0,\ldots,\sigma_{N-1})=&\,f(\sigma_0,\ldots,\sigma_{j-1},-\sigma_j,\sigma_{j+1},\ldots,\sigma_{N-1}),
 \end{align*}
 where $\sigma_0,\ldots,\sigma_{N-1}\in\{1,-1\}$.
 The transfer matrix can now be written as
 \be\label{tmcorrect}
 V=\left(2\sinh2\kx\right)^{\frac{N}{2}} V_y^{\frac12}V_x V_y^{\frac12},
 \eb
 with
 \begin{align}\label{vxvy}
 V_x=\exp\,\biggl\{\kx^*\sum_{j=0}^{N-1}C_j\biggr\},\qquad
 V_y=\exp\,\biggl\{\ky\sum_{j=0}^{N-1}s_js_{j+1}\biggr\}.
 \end{align}
 Here $s_N=\varepsilon_y s_0$ and $\kx^*=\mathrm{arctanh}\,e^{-2\kx}\in\mathbb{R}_{>0}$
  denotes the dual coupling.

 As already mentioned, $V$ commutes with the global spin flip operator
 $U=C_0C_1\ldots C_{N-1}$. It also commutes with the operator $T_{\varepsilon_y}$ of $y$-translations, defined by
 \ben
 \left(T_{\varepsilon_y}f\right)(\sigma_0,\ldots,\sigma_{N-1})=f(\sigma_1,\ldots,\sigma_{N-1},\varepsilon_y\sigma_0),
 \ebn
 as it may be easily checked that
 \ben
 T_{\varepsilon_y}^{\;}s_jT_{\varepsilon_y}^{-1}=s_{j+1},\qquad T_{\varepsilon_y}^{\;}C_jT_{\varepsilon_y}^{-1}=C_{j+1},\qquad j=0,\ldots,N-1,
 \ebn
 with $C_N=C_0$. It is clear that $[T_{\varepsilon_y},U]=0$, and therefore the operators $T_{\varepsilon_y}$, $U$ and $V$   can be diagonalized simultaneously.
 Correlation function
 (\ref{corfun}) with $j_1\leq j_2\leq\ldots\leq j_n$ is equal to the trace ratio
 \begin{align*}
 &\langle\sigma_{j_1,k_1}\ldots\sigma_{j_n,k_n}\rangle=\\
 &=\frac{
 \mathrm{Tr}\left[\left(V^{j_1}T_{\varepsilon_y}^{k_1}s_0T_{\varepsilon_y}^{-k_1}V^{-j_1}\right)\ldots
 \left(V^{j_n}T_{\varepsilon_y}^{k_n}s_0T_{\varepsilon_y}^{-k_n}V^{-j_n}\right){V}^M U^{\frac{1-\varepsilon_x}{2}}\right]}{
 \mathrm{Tr}\left[{V}^M U^{\frac{1-\varepsilon_x}{2}}\right]},
 \end{align*}
 which can be straightforwardly computed in terms of matrix elements of $s_0$ in the
 corresponding basis of normalized common eigenstates.
 \begin{remark}\label{tmwrong}
 The transfer matrix considered in \cite{Hystad,Palmer_Hystad} is given,  in our notation,
 by
 \ben
 V'=\left(2\sinh2\kx\right)^{\frac{N}{2}} V_x^{\frac12}V_y V_x^{\frac12}.
 \ebn
 Since $V'$ is related
 to $V$ in (\ref{tmcorrect}) by a similarity transformation, both definitions work equally well
  if one is interested in the partition function.
 However, $V_x$ does not commute with
 the spin operators $\{s_j\}$, and therefore $V'$ can not be used in the
 calculation of Ising correlation functions. This explains discrepancies reported in
 Section~9 of \cite{Hystad}.

 On the other hand, the transfer matrix $V'$ possesses the same set of eigenvectors
 as the hamiltonian of quantum XY chain in a transverse field.
 Ising spin matrix elements between the eigenvectors of $V'$ should therefore coincide with
 the form factors of $\sigma^x$ between the eigenstates of XY hamiltonian, which were found in \cite{Iorgov5} by the method of separation of variables.
 \end{remark}

 For $j=0,\ldots,N-1$, introduce the operators
 \begin{align*}
 p_j=&\,\underbrace{\sigma_x\otimes\ldots\otimes\sigma_x}_{j\;\text{times}}\otimes
 \,\sigma_z\otimes\mathbf{1}\otimes
 \ldots\otimes\mathbf{1}=C_0\ldots C_{j-1}s_j,\\
 q_j=&\,\underbrace{\sigma_x\otimes\ldots\otimes\sigma_x}_{j\;\text{times}}\otimes
 \,\sigma_y\otimes\mathbf{1}\otimes
 \ldots\otimes\mathbf{1}=iC_0\ldots C_js_j,
 \end{align*}
 satisfying anticommutation relations for the generators of the Clifford algebra:
 \ben
 \left\{p_j,p_k\right\}= \left\{q_j,q_k\right\}=2\delta_{jk},\qquad \left\{p_j,q_k\right\}=0.
 \ebn
 It can be shown that
 \be\label{tmvavp}
 V=\left(2\sinh2\kx\right)^{\frac{N}{2}}\left(\frac{\mathbf{1}+\varepsilon_y U}{2}V_{a}+
 \frac{\mathbf{1}-\varepsilon_y U}{2}V_{p}\right),
 \eb
 where
 \ben
 V_{\nu}=V_{y,\nu}^{\frac12}V_x V_{y,\nu}^{\frac12},\qquad \nu=a,p,
 \ebn
 \ben
 V_x=\exp\biggl\{i\kx^*\sum_{j=0}^{N-1}p_jq_j\biggr\},\qquad
 V_{y,\nu}=\exp\biggl\{-i\ky\sum_{j=0}^{N-1}p_{j+1}q_j\biggr\},
 \ebn
 and $p_N=-p_0$ ($p_N=p_0$) for $\nu=a$ (resp. $\nu=p$). Note
 that $U$ anticommutes with all $\{p_j\}$, $\{q_j\}$ and hence commutes with
 $V_{a,p}$.

 Consider discrete Fourier transforms
 \be\label{Fourier}
 p_{\theta}=\frac{1}{\sqrt{N}}\sum_{j=0}^{N-1}e^{-ij\theta}p_j,\qquad
 q_{\theta}=\frac{1}{\sqrt{N}}\sum_{j=0}^{N-1}e^{-ij\theta}q_j.
 \eb
 Two sets of quasimomenta will be important for us:
 $\boldsymbol{\theta}_{a}=\left\{\frac{\pi}{N},\frac{3\pi}{N},\ldots,2\pi-\frac{\pi}{N}\right\}$
  and  $\boldsymbol{\theta}_{p}=\left\{0,\frac{2\pi}{N},\ldots,2\pi-\frac{2\pi}{N}\right\}$. In both cases
  one has inversion formulas
 \be\label{Fourier_inverse}
 p_j=\frac{1}{\sqrt{N}}\sum_{\theta\in\boldsymbol{\theta}_{\nu}}e^{ij\theta}p_{\theta},\qquad
 q_j=\frac{1}{\sqrt{N}}\sum_{\theta\in\boldsymbol{\theta}_{\nu}}e^{ij\theta}q_{\theta},
 \eb
 and anticommutation relations
 \ben
 \left\{p_{\theta},p_{\theta'}\right\}=\left\{q_{\theta},q_{\theta'}\right\}=
 2\delta_{\theta+\theta',0\;\mathrm{mod}\;2\pi},\qquad \left\{p_{\theta},q_{\theta'}\right\}=0.
 \ebn

 The conjugation by $V_{a,p}$ acts linearly on the generators $\{p_j\}$ and $\{q_j\}$.
 In particular, Fourier components with $\theta\in\boldsymbol{\theta}_{\nu}$
  ($\nu=a,p$) transform as
 \ben
 V_{\nu} \left(\begin{array}{c}
 p_{\theta}  \\
 q_{\theta}
 \end{array}\right)V_{\nu}^{-1}=\left(\begin{array}{cc}
 \cosh\gamma_{\theta} & ie^{-i\theta}w_{\theta} \\
 -ie^{i\theta}w_{-\theta} & \cosh\gamma_{\theta}
 \end{array}\right)
 \left(\begin{array}{c}
 p_{\theta}  \\
 q_{\theta}
 \end{array}\right),
 \ebn
 where the functions $w_{\theta}$ and $\gamma_{\theta}$ are defined by
 \begin{align}
 \label{wtheta}
 w_{\theta}=&\,\cosh^2\kx^*\sinh2\ky\left(1-\alpha e^{i\theta}\right)\left(1-\beta e^{-i\theta}\right),
 \\
 \label{scurve1}
 \cosh\gamma_{\theta}=&\,\cosh2\kx^*\cosh2\ky-\sinh2\kx^*\sinh2\ky\cos\theta,\qquad \gamma_{\theta}\geq0,
 \end{align}
 with $\alpha=\tanh\kx^*\coth\ky$ and $\beta=\tanh\kx^*\tanh\ky$.  It can be easily checked that
 \ben
 \sinh^2\gamma_{\theta}=\cosh^4\kx^*\sinh^22\ky \left(1-\alpha e^{i\theta}\right) \left(1-\alpha e^{-i\theta}\right)
 \left(1-\beta e^{i\theta}\right)\left(1-\beta e^{-i\theta}\right).
 \ebn

 From now on, unless otherwise is stated explicitly, we will work in the ferromagnetic region of parameters. The latter
 is defined by $\kx^*<\ky$ or, equivalently, $\alpha<1$. In this case, it is convenient to write $w_{\pm\theta}$
 in (\ref{wtheta}) as
 \ben
 w_{\pm\theta}=b_{\pm\theta}\sinh\gamma_{\theta},
 \ebn
 with
 \be\label{btheta}
 b_{\theta}=\left[b_{-\theta}\right]^{-1}=\sqrt{\frac{\left(1-\alpha e^{i\theta}\right)\left(1-\beta e^{-i\theta}\right)}{
 \left(1-\beta e^{i\theta}\right)\left(1-\alpha e^{-i\theta}\right)}}.
 \eb
 The branches of square roots in (\ref{btheta}) and below are defined so that they have a positive real part.

 Finally, introduce creation-annihilation operators of $a$- and $p$-fermions, which
 correspond to $\theta\in\boldsymbol{\theta}_a$ and $\theta\in\boldsymbol{\theta}_p$, respectively:
 \begin{align}\label{caops}
 \begin{cases}
 2\psi^{\dag}_{\theta}=e^{-i\theta}\sqrt{b_{\theta}}\,p_{-\theta}-i\sqrt{b_{-\theta}}\,q_{-\theta},\\
 2\psi_{\theta}=\;e^{i\theta}\sqrt{b_{-\theta}}\,p_{\theta}\,+\,i\sqrt{b_{\theta}}\,q_{\theta}.
 \end{cases}
 \end{align}
 Here  $-\theta$ is identified with $2\pi-\theta$ for $\theta\neq0$.
 The operators in each of the two sets satisfy standard anticommutation relations
 \ben
 \{\psi^{\dag}_{\theta},\psi^{\dag}_{\theta'}\}=\{\psi_{\theta},\psi_{\theta'}\}=0,\qquad
 \{\psi^{\dag}_{\theta},\psi_{\theta'}\}=\delta_{\theta,\theta'}.
 \ebn
 Inversion formulas
 \begin{align}\label{inversionfs}
 p_{\theta}=e^{-i\theta}\sqrt{b_{\theta}}\left(\psi^{\dag}_{-\theta}+\psi_{\theta}\right),\qquad
 q_{\theta}=i\sqrt{b_{-\theta}}\left(\psi^{\dag}_{-\theta}-\psi_{\theta}\right),
 \end{align}
 combined with Fourier transforms (\ref{Fourier})--(\ref{Fourier_inverse}),
 allow to express ``antiperiodic''
 creation-annihilation operators as linear combinations of ``periodic'' ones and vice versa.

 These operators transform diagonally under conjugation by  $V_{a,p}$,
 \ben
 V_{\nu} \left(\begin{array}{c}
 \psi^{\dag}_{\theta}  \\
 \psi_{\theta}
 \end{array}\right)V_{\nu}^{-1}=\left(\begin{array}{cc}
 e^{-\gamma_{\theta}} & 0 \\
 0 & e^{\gamma_{\theta}}
 \end{array}\right)
 \left(\begin{array}{c}
 \psi^{\dag}_{\theta}  \\
 \psi_{\theta}
 \end{array}\right),\qquad \theta\in\boldsymbol{\theta}_{\nu}.
 \ebn
 Since $\{s_j\}$, $\{C_j\}$ (as well as $\{p_j\}$, $\{q_j\}$) generate full $2^N\times 2^N$ matrix algebra,
 it follows from Schur's lemma that induced linear transformations fix $V_{a,p}$ up to a scalar multiple.
 Now taking into account that $\mathrm{det}\,V_{a,p}=1$ and $\mathrm{Tr}\,V_{a,p}>0$, we find that
 \ben
 V_{\nu}=\exp\biggl\{-\sum_{\theta\in\boldsymbol{\theta}_{\nu}}\gamma_{\theta}\left(\psi^{\dag}_{\theta}\psi_{\theta}-
 \text{\small$\frac12$}\right)\biggr\},
 \qquad \nu=a,p.
 \ebn
  Fock states
 \be\label{fockstates}
 |\theta_1,\ldots,\theta_k\rangle_{\nu}=\psi^{\dag}_{\theta_1}\ldots \psi^{\dag}_{\theta_k}|vac\rangle_{\nu},
 \qquad \theta_1,\ldots,\theta_k\in\boldsymbol{\theta}_{\nu},
 \eb
 are therefore right eigenvectors of $V_{\nu}$, with eigenvalues
 equal to
 \ben \exp\left\{\frac12\sum_{\theta\in\boldsymbol{\theta}_{\nu}}\gamma_{\theta}-
 \sum_{i=1}^k\gamma_{\theta_i}\right\}.
 \ebn
 Here $|vac\rangle_{\nu}$ is a vector annihilated by all $\psi_{\theta}$ with
 $\theta\in\boldsymbol{\theta}_{\nu}$.

  The operators $\{s_j\}$, $\{C_j\}$,
 $\{p_j\}$, $\{q_j\}$ are hermitian, hence $p^{\dag}_{\theta}=p_{-\theta}$, $q^{\dag}_{\theta}=q_{-\theta}$
 and the operator $\psi^{\dag}_{\theta}$ is hermitian conjugate of $\psi_{\theta}$, just as the notation
 suggests. Introduce the dual vacua $_{\nu}\langle vac|=|vac\rangle_{\nu}^{\dag}$ annihilated by all $\psid_{\theta}$ with $\theta\in\boldsymbol{\theta}_{\nu}$. Left eigenvectors
 of $V_{\nu}$ can then be written as
 \ben
 _{\nu}\langle \theta_k,\ldots,\theta_1|=|\theta_1,\ldots,\theta_k\rangle_{\nu}^{\dag}=\,_{\nu}\langle vac|\psi_{\theta_k}\ldots \psi_{\theta_1},\qquad \theta_1,\ldots,\theta_k\in\boldsymbol{\theta}_{\nu}.
 \ebn
 We will impose the normalization condition $_{\nu}\langle vac|vac\rangle_{\nu}=1$, which
 automatically implies  orthonormality of the states (\ref{fockstates}).

 Recall that only a subset of eigenvectors of $V_{a,p}$ are actual eigenstates of
 $V$, cf. (\ref{tmvavp}). For $\varepsilon_y=1$ (i.e. periodic boundary conditions on spin variables)
 only the states with even number of particles are eigenvectors of the full transfer
 matrix. For $\varepsilon_y=-1$, the number of particles in the surviving
 states of both types is odd.
  This follows e.g. from an explicit diagonalization of
 $V_{a,p}$ in the limit $\kx^*\rightarrow0$, and observation that the eigenvectors
 for $\kx^*>0$ are related to those with $\kx^*=0$ by products of two-mode Bogoliubov
 transformations, which commute with $U$. In fact since $U$  also anticommutes with $\{\psi^{\dag}_{\theta}\}$,
 $\{\psi_{\theta}\}$, it is sufficient to show that $\mathbb{Z}_2$-charges of the vacuum vectors
 $|vac\rangle_{a}$ and $|vac\rangle_{p}$ are equal to $+1$ and $-1$, respectively.

  To avoid confusion, in what follows we reserve the letters $\psi^{\dag}_{\theta}$, $\psi_{\theta}$
 for the creation-annihilation operators with $\theta\in\boldsymbol{\theta}_a$ and denote
 the corresponding operators with $\theta\in\boldsymbol{\theta}_p$ by $\varphi^{\dag}_{\theta}$,
 $\varphi_{\theta}$.
 Our aim is to compute spin form factors
 \be\label{spinff}
  \mathcal{F}_{m,n}^{(l)}\left(\boldsymbol{\theta},\boldsymbol{\theta}'\right)=
  {}_a\langle \theta_1,\ldots,\theta_m|s_l|\theta_1',\ldots,\theta_n'\rangle_p=
  \left[\bigl.{}_p\langle \theta'_n,\ldots,\theta'_1|s_l|\theta_m,\ldots,\theta_1\rangle_a\right]^*,
 \eb
 where $\theta_1,\ldots,\theta_m\in\boldsymbol{\theta}_a$, $\theta_1',\ldots,\theta_n'\in\boldsymbol{\theta}_p$ and $*$  in the last expression
 denotes complex conjugation. As explained above, matrix elements (\ref{spinff}) are non-zero only if
 $m$ and $n$ are simultaneously even or odd.
 These are the only nontrivial form factors: since $s_l$ anticommutes with $U$,
 its matrix elements between the eigenstates of $V$ of the same type vanish.

 \begin{remark}  Let us show that the eigenstates
 of $V$ constructed above diagonalize the translation operator $T_{\varepsilon_y}$
 (since $\gamma_{\theta}= \gamma_{-\theta}$, the eigenvalues of $V$ are degenerate and this is not automatic). Though translation invariance will be manifest
 in the final answer, which satisfies
 \be\label{trinv}
 \mathcal{F}_{m,n}^{(l)}\left(\boldsymbol{\theta},\boldsymbol{\theta'}\right)=
 e^{il\left(\sum_{i=1}^n\theta'_i-\sum_{i=1}^m\theta_i\right)}
 \mathcal{F}_{m,n}^{(0)}\left(\boldsymbol{\theta},\boldsymbol{\theta'}\right),
 \eb
 it is instructive to explain it already at this stage.

 Indeed, the vacua $|vac\rangle_{a,p}$ are
 non-degenerate eigenstates of the periodic Ising transfer matrix, and therefore they are
 also eigenvectors of $T_+$. It can be deduced from the limit $\kx^*\rightarrow0$ by continuity
 that the corresponding eigenvalues are both equal to $1$. It is also straightforward to check
 that
 \ben
 T^{\,}_{\pm}\left(\begin{array}{c}p_j \\ q_j\end{array}\right)T^{-1}_{\mp}\frac{\mathbf{1}\pm\epsilon U}{2}=
 \left(\begin{array}{c}p_{j+1} \\ q_{j+1}\end{array}\right)\frac{\mathbf{1}\pm\epsilon U}{2},
 \qquad j=0,\ldots,N-1,
 \ebn
 where $\epsilon=\pm1$  and $p_N=\epsilon p_0$, $q_N=\epsilon q_0$. Performing discrete
 Fourier transforms, we obtain
 \begin{align}
 \label{trans1}
 T^{\,}_{\pm}\left(\begin{array}{c}\psi^{\dag}_{\theta} \\ \psi_{\theta} \end{array}\right)T^{-1}_{\mp}\frac{\mathbf{1}\mp U}{2}=&\,
 \left(\begin{array}{c}e^{-i\theta}\psi^{\dag}_{\theta} \\ \,e^{\,i\theta}\,\psi_{\theta} \end{array}\right)\frac{\mathbf{1}\mp U}{2},\\
 \label{trans2}T^{\,}_{\pm}\left(\begin{array}{c}\varphi^{\dag}_{\theta} \\ \varphi_{\theta} \end{array}\right)T^{-1}_{\mp}\frac{\mathbf{1}\pm U}{2}=&\,
 \left(\begin{array}{c}e^{-i\theta}\varphi^{\dag}_{\theta} \\ \,e^{\,i\theta}\,\varphi_{\theta} \end{array}\right)\frac{\mathbf{1}\pm U}{2},
 \end{align}
 which in turn yields the desired formulas:
 \begin{align}
 \label{trans3}
  T_+|\theta_1,\ldots,\theta_{2k}\rangle_{a,p}\quad=&\,\,
 e^{-i\sum_{i=1}^{2k}\theta_i}\;|\theta_1,\ldots,\theta_{2k}\rangle_{a,p},\\
 \label{trans4}
 T_-|\theta_1,\ldots,\theta_{2k+1}\rangle_{a,p}=&\,
 e^{-i\sum_{i=1}^{2k+1}\theta_i}|\theta_1,\ldots,\theta_{2k+1}\rangle_{a,p}.
 \end{align}
 \end{remark}

 \section{Spin form factors and induced rotations\label{sec_rotations}}
 Just as in the above case of $V_{a,p}$, conjugation by the spin operator $s_{l}$
 acts linearly on the Clifford algebra generators $\{p_j\}$,
 $\{q_j\}$:
 \be\label{indtr}
 \begin{cases}
 s_lp_js_l^{-1}=\mathrm{sgn}(l-j)p_j,\\  s_lq_js_l^{-1}=\mathrm{sgn}(l-1-j)q_j,
 \end{cases}\qquad
 j=0,\ldots,N-1,
 \eb
 where $\mathrm{sgn}(x)=1$ if $x\geq0$ and $-1$ if $x<0$. Recall that induced linear
 transformations fix $s_l$ up to a scalar multiple.

 Let us combine
 $\{\psi^{\dag}_{\theta}\}$, $\{\psi^{\,}_{\theta}\}$ and  $\{\varphi^{\dag}_{\theta}\}$, $\{\varphi^{\,}_{\theta}\}$
 into $N$-dimensional column vectors $\vec{\psi}^{\,\dag}$, $\vec{\psi}$, ${\vec{\varphi}}^{\,\dag}$, $\vec{\varphi}$.
 Their entries are thus given by $2^N\times 2^N$ matrices. It will be very convenient to write the result
 of conjugation of $\{\varphi^{\dag}_{\theta}\}$, $\{\varphi^{\,}_{\theta}\}$ by $s_l$
 as a linear combination of $\{\psi^{\dag}_{\theta}\}$, $\{\psi^{\,}_{\theta}\}$:
 \be\label{abcd2}
   s_l\left(\begin{array}{l}\vec{\varphi}^{\,\dag} \\ \vec{\varphi} \end{array}\right)
  s_l^{-1} =
  \left(\begin{array}{cc} A & B \\ C & D\end{array}\right)
  \left(\begin{array}{l}\vec{\psi}^{\,\dag} \\ \vec{\psi} \end{array}\right),
 \eb
 where $A$, $B$, $C$, $D$ are $N\times N$ matrices.
 The entries of the $2N$-column in the l.h.s. satisfy canonical anticommutation relations,
 from which it can be deduced that
 \be\label{abcdrels1}
 AB^T+BA^T=CD^T+DC^T=0,\qquad AD^T+BC^T=\mathbf{1}.
 \eb
 Since $s^{\,}_l=s_l^{-1}=s_l^{\dag}$, the operator $s_l$ is unitary. This yields further
 relations
 \be\label{abcdrels2}
 \bar{A}=D,\qquad \bar{B}=C.
 \eb

 Vacuum vectors $|vac\rangle_{\nu}$ ($\nu=a,p$) are fixed by the annihilation operators $\{\psi_{\theta}\}$,
 $\{\varphi_{\theta}\}$ and the normalization $_{\nu}\langle vac|vac\rangle_{\nu}=1$
 up to inessential pure phase factors. This can be used to establish a relation between
 the two vacua.
 \begin{lemma}\label{lempvac}
 We have
 \be\label{pvac}
 |vac\rangle_p=\lambda\cdot s_l^{-1} \mathcal{O}|vac\rangle_a,
 \eb
 where
 \begin{align*}
 |\lambda|=|\mathrm{det}\,D|^{\frac12},\qquad \mathcal{O}=\exp\biggl\{-\frac12
 \sum_{\theta,\theta'\in\boldsymbol{\theta}_a}\psi^{\dag}_{\theta}\left(D^{-1}C\right)_{\theta,\theta'}
 \psi^{\dag}_{\theta'}\biggr\}.
 \end{align*}
 \end{lemma}
 \pf Lemma is formulated and will be proved under assumption that $D$ is invertible. This point will be checked
 in Section~\ref{sec_cauchy}, where $\mathrm{det}\,D$ is shown to be non-zero by explicit computation.

 Acting on $ s_l^{-1} \mathcal{O}|vac\rangle_a$ with the $N$-column $\vec{\varphi}$ of annihilation operators, one finds
 \be\label{vacaux}
 \vec{\varphi}s_l^{-1} \mathcal{O}|vac\rangle_a=s_l^{-1} (C\vec{\psi}^{\,\dag}+D\vec{\psi})\mathcal{O}|vac\rangle_a=
 s_l^{-1} \mathcal{O}(C\vec{\psi}^{\,\dag}+D\mathcal{O}^{-1}\vec{\psi}\mathcal{O})|vac\rangle_a.
 \eb
 For two matrices $X$, $Y$ satisfying $[X,[X,Y]]=0$ one has $e^XYe^{-X}=Y+[X,Y]$. From this
 and the fact that $D^{-1}C$ is skew-symmetric because of (\ref{abcdrels1}) we get
 \be\label{opsio}
 \mathcal{O}^{-1}\vec{\psi}\mathcal{O}=-D^{-1}C\vec{\psi}^{\,\dag}+\vec{\psi}.
 \eb
 Since $\vec{\psi}|vac\rangle_a=0$, it follows from (\ref{vacaux}) and (\ref{opsio}) that $\vec{\varphi}s_l^{-1} \mathcal{O}|vac\rangle_a=0$, and thus
 the vector $s_l^{-1} \mathcal{O}|vac\rangle_a$ is proportional to $|vac\rangle_p$.

 Absolute value of the scalar multiple $\lambda$ in (\ref{pvac}) can be determined from the normalization.
 Indeed, the unitarity of $s_l$ implies that
 \be\label{vevd}
 1={}_p\langle vac|vac\rangle_p=|\lambda|^2{}_a\langle vac |\mathcal{O}^{\dag}\mathcal{O}|vac\rangle_a.
 \eb
 To compute the vacuum expectation value on the right, one can e.g. write $D^{-1}C$ as $G^TFG$,
 where $G$ is unitary and
 $F$ has block-diagonal form, with $2\times2$ skew-symmetric blocks containing
 nontrivial eigenvalues of $D^{-1}C$ and $1\times1$ blocks containing zeros.
 Introducing new creation-annihilation operators
 $\vec{\xi}^{\,\dag}=G\vec{\psi}^{\,\dag}$, $\vec{\xi}=\bar{G}\vec{\psi}$ which satisfy
 canonical anticommutation relations, and using that $|vac\rangle_a$ is annihilated
 by $\vec{\xi}$, one obtains
 \begin{align*}
 {}_a\langle vac |\mathcal{O}^{\dag}\mathcal{O}|vac\rangle_a &\,=\left[\mathrm{det}\left(
 \mathbf{1}+FF^{\dag}\right)\right]^{\frac12}=\\ &\,=
 \left[\mathrm{det}
 \left(\mathbf{1}+\left(D^{-1}C\right)\left(D^{-1}C\right)^{\dag}\right)\right]^{\frac12}
 =|\mathrm{det}\,D|^{-1}.
 \end{align*}
 The last equality follows from the relation
 $CC^{\dag}=\mathbf{1}-DD^{\dag}$, which can be derived from (\ref{abcdrels1})--(\ref{abcdrels2}).
 Now (\ref{vevd}) implies that $|\lambda|=|\mathrm{det}\,D|^{\frac12}$ and the lemma is proved.
 \epf

 Since $T_+|vac\rangle_{a,p}=|vac\rangle_{a,p}$, the factor $\lambda={}_a\langle vac|s_l|vac\rangle_p$ is  independent of $l$
 and thus its phase can always be absorbed into the definition of vacua.
 We will therefore assume below that $\lambda\in\Rb_{>0}$. Under this convention,
 it follows from Lemma~\ref{lempvac} that
 \be\label{vevsigma}
 {}_a\langle vac|s_{l}|vac\rangle_p=|\mathrm{det}\,D|^{\frac12}.
 \eb
 Note that  in the limit $N\rightarrow\infty$ the l.h.s. of this formula
 gives Ising spontaneous magnetization.

 Let us now compute general form factors $\mathcal{F}_{m,n}^{(l)}(\boldsymbol{\theta},\boldsymbol{\theta}')$.
 By (\ref{spinff}) and (\ref{pvac}), one has
 \begin{align*}
 \lambda^{-1}\mathcal{F}_{m,n}^{(l)}(\boldsymbol{\theta},\boldsymbol{\theta}')=
 {}_a\langle vac|\psi^{\,}_{\theta_1}\ldots\psi^{\,}_{\theta_m}s_l
 \varphi^{\dag}_{\theta'_1}\ldots\varphi^{\dag}_{\theta'_n}s_l^{-1}\mathcal{O}|vac\rangle_a=\\
 ={}_a\langle vac|\psi^{\,}_{\theta_1}\ldots\psi^{\,}_{\theta_m}
 (A\vec{\psi}^{\,\dag}+B\vec{\psi})_{\theta'_1}\ldots(A\vec{\psi}^{\,\dag}+B\vec{\psi})_{\theta'_n}\mathcal{O}|vac\rangle_a,
 \end{align*}
 where we used (\ref{abcd2}) to pass from the first to the second line. Now pull $\mathcal{O}$ in the last
 expression to the left vacuum, on which it acts as the identity operator. This can be done using (\ref{opsio})
 and the fact that $\mathcal{O}$ commutes with all entries of $\vec{\psi}^{\,\dag}$. Moreover, (\ref{abcdrels1}) implies
 that
 \ben
 \mathcal{O}^{-1}(A\vec{\psi}^{\,\dag}+B\vec{\psi})\mathcal{O}=(A-BD^{-1}C)\vec{\psi}^{\,\dag}+B\vec{\psi}
 =D^{-T}\vec{\psi}^{\,\dag}+B\vec{\psi},
 \ebn
 where $D^{-T}=\left(D^{-1}\right)^T$, so that
 \begin{align*}
 \lambda^{-1}\mathcal{F}_{m,n}^{(l)}(\boldsymbol{\theta},\boldsymbol{\theta}')={}
 {}_a\langle vac|(-D^{-1}C\vec{\psi}^{\,\dag}+\vec{\psi})_{\theta_1}\ldots(-D^{-1}C\vec{\psi}^{\,\dag}+\vec{\psi})_{\theta_m}\times&\\
 \times(D^{-T}\vec{\psi}^{\,\dag}+B\vec{\psi})_{\theta'_1}\ldots(D^{-T}\vec{\psi}^{\,\dag}+B\vec{\psi})_{\theta'_n}|vac\rangle_a.&
 \end{align*}
 Vacuum expectation value of any product of linear combinations of $\{\psi^{\dag}_{\theta}\}$, $\{\psi^{\,}_{\theta}\}$ can
 be easily computed using Wick theorem. The result is given by the pfaffian of the matrix of pairings between
 different combinations. In our case, there are three types of pairings:
 \begin{align*}
 {}_a\langle vac|(-D^{-1}C\vec{\psi}^{\,\dag}+\vec{\psi})_{\theta_j}(-D^{-1}C\vec{\psi}^{\,\dag}+\vec{\psi})_{\theta_k}|vac\rangle_a=
 &\,\left(D^{-1}C\right)_{\theta_j,\theta_k},\\
  {}_a\langle vac|(-D^{-1}C\vec{\psi}^{\,\dag}+\vec{\psi})_{\theta_j}(D^{-T}\vec{\psi}^{\,\dag}+B\vec{\psi})_{\theta'_k}|vac\rangle_a=
 &\,D^{-1}_{\theta_j,\theta'_k},\\
  {}_a\langle vac|(D^{-T}\vec{\psi}^{\,\dag}+B\vec{\psi})_{\theta'_j}(D^{-T}\vec{\psi}^{\,\dag}+B\vec{\psi})_{\theta'_k}|vac\rangle_a=
 &\,\left(BD^{-1}\right)_{\theta'_j,\theta'_k}.
 \end{align*}

 Accordingly, we obtain
 \begin{lemma}\label{mpff}
  Form factors $\mathcal{F}_{m,n}^{(l)}(\boldsymbol{\theta},\boldsymbol{\theta}')$ of Ising spin have the
 following representation in terms
 of induced rotations (\ref{abcd2}):
 \begin{align}
 \label{ffpfaffian}
 \mathcal{F}_{m,n}^{(l)}(\boldsymbol{\theta},\boldsymbol{\theta}')=
 |\mathrm{det}\,D|^{\frac12}\cdot\mathrm{Pf}\,R,
 \qquad R=\left(
 \begin{array}{cc}
 R_{\boldsymbol{\theta}\times\boldsymbol{\theta}} &
 R_{\boldsymbol{\theta}\times\boldsymbol{\theta}'} \\
 R_{\boldsymbol{\theta}'\times\boldsymbol{\theta}} &
 R_{\boldsymbol{\theta}'\times\boldsymbol{\theta}'}
 \end{array}\right),
 \end{align}
 where matrix elements of the blocks of skew-symmetric $(m+n)\times(m+n)$ matrix $R$ are given by
 \begin{align}
 \label{r11}\left(R_{\boldsymbol{\theta}\times\boldsymbol{\theta}}\right)_{jk}=\left(D^{-1}C\right)_{\theta_j,\theta_k},
 &\qquad j,k=1,\ldots,m,\\
 \label{r12} \left(R_{\boldsymbol{\theta}\times\boldsymbol{\theta}'}\right)_{jk}=
 -\left(R_{\boldsymbol{\theta}'\times\boldsymbol{\theta}}\right)_{kj}=D^{-1}_{\theta_j,\theta'_k},
 &\qquad j=1,\ldots,m,\quad k=1,\ldots,n,\\
 \label{r22}\left(R_{\boldsymbol{\theta}'\times\boldsymbol{\theta}'}\right)_{jk}=
 \left(BD^{-1}\right)_{\theta'_j,\theta'_k},
 &\qquad j,k=1,\ldots,n.
 \end{align}
 \end{lemma}
 \noindent Observe that
 matrix elements (\ref{r11})--(\ref{r22}) coincide with normalized two-particle form factors:
 \begin{align}\label{2pffmatr1}
 \frac{_a\langle \theta,\theta'|s_l|vac\rangle_p}{_a\langle vac|s_l|vac\rangle_p}=&\,
 \left(D^{-1}C\right)_{\theta,\theta'},\\
 \label{2pffmatr2}
 \frac{_a\langle \theta|s_l|\theta'\rangle_p}{_a\langle vac|s_l|vac\rangle_p}=&\,
 D^{-1}_{\theta,\theta'},\\
 \label{2pffmatr3} \frac{_a\langle vac|s_l|\theta,\theta'\rangle_p}{_a\langle vac|s_l|vac\rangle_p}=&\,
 \left(BD^{-1}\right)_{\theta,\theta'}.
 \end{align}

 The reader might have noticed that the only important point for the above derivation
 is the formula (\ref{abcd2}). In particular, we have not as yet used the fact that $s_l$ belongs
 to the Clifford group, i.e. that $\vec{\psi}^{\,\dag}$, $\vec{\psi}$ and $\vec{\varphi}^{\,\dag}$, $\vec{\varphi}$ are related
 by a linear transformation. However, this does play a role in the proof of the following Lemma,
 which gives the explicit form of the induced rotation matrix.

 \begin{lemma}\label{lemabcd}
 Matrix elements of $A$, $B$, $C$, $D$ are given by
 \begin{align}
 \label{elemsad}\bar{A}_{\theta,\theta'}=D_{\theta,\theta'}=&\,
\frac{e^{-i(l-\frac12)(\theta-\theta')}}{2iN\sin\frac{\theta'-\theta}{2}}
\left(\sqrt{\frac{b_{\theta'}}{b_{\theta}}}+\sqrt{\frac{b_{\theta}}{b_{\theta'}}}\right),\\
 \label{elemsbc}\bar{B}_{\theta,\theta'}=C_{\theta,\theta'}=&\,
\frac{e^{-i(l-\frac12)(\theta+\theta')}}{2iN\sin\frac{\theta+\theta'}{2}}
\left(\sqrt{b_{\theta}b_{\theta'}}-
\frac{1}{\sqrt{b_{\theta}b_{\theta'}}}
\right),
 \end{align}
 where $\theta\in\boldsymbol{\theta}_p$, $\theta'\in\boldsymbol{\theta}_a$.
 \end{lemma}
 \pf  Express  $\varphi_{\theta}$ in terms of $p_{j}$, $q_{j}$ using (\ref{caops}) and the Fourier
 transform (\ref{Fourier}), compute the result of conjugation by $s_l$ using (\ref{indtr}), and rewrite
 it in terms of $\{\psi^{\dag}_{\theta'}\}$, $\{\psi_{\theta'}\}$ using the inverse transform (\ref{Fourier_inverse})
 and the formula (\ref{inversionfs}). Explicitly,
 \begin{align*}
 &2s_l^{\,}\varphi_{\theta}s_l^{-1}=s_l^{\,}\left(e^{i\theta}\sqrt{b_{-\theta}}\,p_{\theta}+
 i\sqrt{b_{\theta}}\,q_{\theta}\right)s_l^{-1}=\\
 &=\frac{1}{\sqrt{N}}\sum_{j=0}^{N-1}e^{-ij\theta}
 \left(e^{i\theta}\sqrt{b_{-\theta}}\,s^{\,}_l p_js^{-1}_l+
 i\sqrt{b_{\theta}}\,s^{\,}_l q_js^{-1}_l\right)=\\
 &=\frac{1}{\sqrt{N}}\sum_{j=0}^{N-1}e^{-ij\theta}\left(e^{i\theta}\sqrt{b_{-\theta}}\,\mathrm{sgn}(l-j)p_j+
 i\sqrt{b_{\theta}}\,\mathrm{sgn}(l-1-j)q_j\right)=\\
 &=\frac{1}{N}\sum_{\theta'\in\boldsymbol{\theta}_a}\sum_{j=0}^{N-1}
 e^{ij(\theta'-\theta)}\left(e^{i\theta}\sqrt{b_{-\theta}}\,\mathrm{sgn}(l-j)p_{\theta'}+
 i\sqrt{b_{\theta}}\,\mathrm{sgn}(l-1-j)q_{\theta'}\right)=\\
 &=-\frac{2}{N}\sum_{\theta'\in\boldsymbol{\theta}_a}\frac{e^{il(\theta'-\theta)}}{1-e^{i(\theta'-\theta)}}
 \left(e^{i\theta'}\sqrt{b_{-\theta}}\,p_{\theta'}+
 i\sqrt{b_{\theta}}\,q_{\theta'}\right)=\\
 &=-\frac{2}{N}\sum_{\theta'\in\boldsymbol{\theta}_a}\frac{e^{il(\theta'-\theta)}}{1-e^{i(\theta'-\theta)}}
 \left\{\left(\sqrt{\frac{b_{\theta'}}{b_{\theta}}}-\sqrt{\frac{b_{\theta}}{b_{\theta'}}}\right)\psi^{\dag}_{-\theta'}+
 \left(\sqrt{\frac{b_{\theta'}}{b_{\theta}}}+\sqrt{\frac{b_{\theta}}{b_{\theta'}}}\right)\psi_{\theta'}\right\},
 \end{align*}
 which gives $C$ and $D$. The formulas for $A$ and $B$ immediately follow from (\ref{abcdrels2}).
 \epf
 \begin{remark}
 An essentially equivalent result was obtained in \cite{Hystad}, Theo\-rem 4.1. However, the function $b_{\theta}$
 appearing in the kernels of $A$, $B$, $C$, $D$ differs from the one  in \cite{Hystad}. This seems to be related to the
 inappropriate initial
 choice of the transfer matrix, see Remark~\ref{tmwrong}.
 \end{remark}

 \section{Elliptic parametrization of the Ising spectral curve\label{elpar}}
 Let us now go back to the lattice dispersion relation (\ref{scurve1}). Introduce the variables
 $z=e^{i\theta}$, $\lambda=e^{\gamma_{\theta}}$ and rewrite it as an algebraic curve
 \be\label{scurve2}
 \sinh2\kx \frac{\lambda+\lambda^{-1}}{2}+\sinh2\ky\frac{z+z^{-1}}{2}=\cosh2\kx\cosh2\ky.
 \eb
 Topologically this is a torus realized as
 a two-fold covering of $\mathbb{P}^1$, branched at 4 points
 \begin{align*}
 z_{1,2}&\,=\left(\tanh\kx^*\coth\ky\right)^{\pm1}=\alpha^{\pm1},\\
 z_{3,4}&\,=\left(\tanh\kx^*\tanh\ky\right)^{\pm1}=\beta^{\pm1}.
 \end{align*}
 The points $\beta<\alpha<\alpha^{-1}<\beta^{-1}$ can be mapped to $0<k<k^{-1}<\infty$ by a M\"obius transformation.
 Such transformations preserve the anharmonic ratio $\displaystyle(z_1,z_2;z_3,z_4)=\frac{(z_1-z_3)(z_2-z_4)}{(z_2-z_3)(z_1-z_4)}$, and therefore
 \ben
 k=\frac{\beta-\alpha}{\alpha\beta-1}=\frac{\sinh2\kx^*}{\sinh2\ky}.
 \ebn

 The parameter $k$ plays the role of the modulus of the Jacobi elliptic functions uniformizing  (\ref{scurve2}). One also needs to introduce elliptic integrals
 \ben
 K=\int_0^1\frac{dt}{\sqrt{(1-t^2)(1-k^2t^2)}},\qquad K'=\int_0^1\frac{dt}{\sqrt{(1-t^2)(1-k'^2t^2)}}.
 \ebn
 where the complementary modulus is defined by $k'^2=1-k^2$. It is known (see Theorem~2.2.1 in \cite{Palmer_book})
 that the functions
 \begin{align}
 \label{zu}z(u)&\,=\frac{\sn\left(u+i\eta\right)}{\sn\left(u-i\eta\right)},\\
 \label{lu}\lambda(u)&\,=\left[k\,\sn\left(u+i\eta\right)\sn\left(u-i\eta\right)\right]^{-1},
 \end{align}
 with $u\in[-K,K)\times i[-K',K')$, provide a uniformization of the Ising spectral curve (\ref{scurve2}). The real parameter $\eta\in\left(-\frac{K'}{2},0\right)$ is determined by $\sinh2\kx=i\,\sn\,2i\eta$
 (it is related to $a$ of \cite{Palmer_book} by $a=\eta+\frac{K'}{2}$).

 Elliptic parametrization (\ref{zu})--(\ref{lu}) establishes a one-to-one correspondence between the ``physical'' cycle $\mathcal{C}_{\theta}=\left\{(z,\lambda)=(e^{i\theta},e^{\gamma_{\theta}})\,|\,\theta\in[0,2\pi)\right\}$
 and the set
 $\mathcal{C}_u=\left\{u\,|\,\mathrm{Re}\,u\in[-K,K),\,\mathrm{Im}\,u=0\right\}$.
 It is convenient to denote by $u_{\theta}$ the image of the point  $(e^{i\theta},e^{\gamma_{\theta}})\in \mathcal{C}_{\theta}$ in $\mathcal{C}_{u}$. In particular, one has $u_0=-K$, $u_{\pi}=0$ and, more generally, $u_{\theta}=-u_{2\pi-\theta}$ for $\theta\in(0,\pi]$.

  The rest of this section is devoted to the
  derivation of a suitable elliptic representation for the elements of the induced rotation matrix.
  We proceed by establishing a number of identities between the two parametrizations. First
   note that for $\theta\in[0,2\pi)$ holds
  \be\label{pmgq2}
  e^{-\frac{\gamma_{\theta}\pm i\theta}{2}}=-\sqrt{k}\;\sn(u_{\theta}\mp i\eta).
  \eb
  Indeed, the squares of both sides of this relation coincide because of (\ref{zu})--(\ref{lu}).
  Since they never vanish, it is sufficient to fix the sign for one value of $\theta$. Set
  $\theta=\pi$, then the l.h.s. becomes $\mp ie^{-\gamma_{\pi}/2}$ and $\sn(u_{\theta}\mp i\eta)$ in the r.h.s.
  reduces to $\mp\sn\,i\eta$. On the other hand $i\,\sn\,i\eta>0$, and the result follows.

  Setting $u=u_{\theta}$, $u'=u_{\theta'}$, $v=i\eta$ in the easily verified formula
  \ben
  \frac{\sn(u+v)\,\sn(u'-v)-\sn(u-v)\,\sn(u'+v)}{1-k^2\sn(u+v)\,\sn(u-v)\,\sn(u'+v)\sn(u'-v)}=
  -\sn(u-u')\,\sn\,2v,
  \ebn
  and using (\ref{pmgq2}), one finds that
  \be\label{snttp}
  \sn(u_{\theta}-u_{\theta'})=\sinh2\ky\;\frac{\sin\frac{\theta-\theta'}{2}}{
  \sinh\frac{\gamma_{\theta}+\gamma_{\theta'}}{2}}.
  \eb
  This identity has several important consequences. Notice especially that
  for $\theta'=0,\pi,2\pi-\theta$ it
  reduces to
  \begin{align}
  \label{snaux1a}
  \frac{\sn(u_{\theta}+K)}{\sinh2\ky}=&\,\frac{\sin\frac{\theta}{2}}{
  \sinh\frac{\gamma_{\theta}+\gamma_{0}}{2}},\\
  \label{snaux1b}\frac{\sn\,u_{\theta}}{\sinh2\ky}=-&\,\frac{\cos\frac{\theta}{2}}{
  \sinh\frac{\gamma_{\pi}+\gamma_{\theta}}{2}},\\
  \label{snaux1c}\frac{\sn\,2u_{\theta}}{\sinh2\ky}=
  -&\,\frac{\sin\theta}{\sinh\gamma_{\theta}}.
  \end{align}
  Since $\cosh\gamma_{\theta}-\cosh\gamma_{\theta'}=\sinh2\kx^*\sinh2\ky(\cos\theta'-\cos\theta)$
  by (\ref{scurve1}),
  it can be inferred from (\ref{snaux1b}) that
  \begin{align}
  \label{snaux2}k\;\sn^2u_{\theta}=&\,\frac{\sinh\frac{\gamma_{\pi}-
  \gamma_{\theta}}{2}}{\sinh\frac{\gamma_{\pi}+\gamma_{\theta}}{2}},\\
  \label{snaux4}1-k^2\,\sn^2u_{\theta}\,\sn^2u_{\theta'}=&\,\frac{\sinh\gamma_{\pi}
  \sinh\frac{\gamma_{\theta}+\gamma_{\theta'}}{2}}{
  \sinh\frac{\gamma_{\pi}+\gamma_{\theta}}{2}\sinh\frac{\gamma_{\pi}+\gamma_{\theta'}}{2}}.
  \end{align}

    Next we want to compute $\cn\,u_{\theta}$ and $\dn\,u_{\theta}$. For that, set $\theta'=\theta$
  in the last identity and combine the result with the doubling formula
  $\displaystyle\sn\,2u=\frac{2\,\sn\,u\,\cn\,u\,\dn\,u}{1-k^2\sn^4u}$ and  (\ref{snaux1b}) to fix the product $\cn\,u_{\theta}\,\dn\,u_{\theta}$. Together with (\ref{snaux1a}) (since $\displaystyle\sn(u+K)=\frac{\cn\,u}{\dn\,u}$ and
  $\cn\,u_{\theta},\,\dn\,u_{\theta}\,\geq 0$),
  this gives
  \begin{align}
  \label{sndn}
  \cn\,u_{\theta}=\sin\frac{\theta}{2}\;\sqrt{\frac{\sinh2\ky\,\sinh\gamma_{\pi}}{
  \sinh\frac{\gamma_{\theta}+\gamma_{0}}{2}\sinh\frac{\gamma_{\theta}+\gamma_{\pi}}{2}}}\,,\quad
  \dn\,u_{\theta}=\sqrt{\frac{\sinh\gamma_{\pi}\sinh\frac{\gamma_{\theta}+\gamma_{0}}{2}}{
  \sinh2\ky\sinh\frac{\gamma_{\theta}+\gamma_{\pi}}{2}}}\,.
  \end{align}

 To find an elliptic representation for $b_{\theta}$ in (\ref{btheta}), recall that
 \ben
 b_{\theta}=\cosh^2\kx^*\sinh2\ky\,\frac{\left(1-\alpha e^{i\theta}\right)\left(1-\beta e^{-i\theta}\right)}{\sinh\gamma_{\theta}}.
 \ebn
 As a function of $\theta$, the numerator of this formula is a linear
 combination of $1$, $\cos\theta$ and $\sin\theta$.
 On the other hand, (\ref{snaux2}) and (\ref{snaux4}) imply that
 \begin{align}
 \label{snaux5}\frac{2k\;\sn^2u_{\theta}}{1-k^2\sn^4u_{\theta}}=&\,\frac{\cosh\gamma_{\pi}-\cosh\gamma_{\theta}}{\sinh\gamma_{\pi}\sinh\gamma_{\theta}},
 \\
 \label{snaux6}\frac{1+k^2\sn^4u_{\theta}}{1-k^2\sn^4u_{\theta}}=&\,\frac{\cosh\gamma_{\pi}\cosh\gamma_{\theta}-1}{\sinh\gamma_{\pi}\sinh\gamma_{\theta}}.
 \end{align}
 Hence $b_{\theta}$ is a linear combination of the l.h.s.'s of (\ref{snaux5}), (\ref{snaux6})
 and (\ref{snaux1c}). A little calculation then gives
 \ben\label{ellbtheta1}
 b_{\theta}=ik\,\sn\,2u_{\theta}-\frac{2k^2\sn^2u_{\theta}}{1-k^2\sn^4u_{\theta}}+\frac{1+k^2\sn^4u_{\theta}}{1-k^2\sn^4u_{\theta}}
 =\frac{\left(\dn\,u_{\theta}+ik\,\sn\,u_{\theta}\,\cn\,u_{\theta}\right)^2
 }{1-k^2\sn^4u_{\theta}}.
 \ebn
 Taking the square root and fixing the signs via $\sqrt{b_{\pi}}=1$, one obtains
 \be\label{ellbtheta2}
 \left[\sqrt{b_{\theta}}\,\right]^{\pm1}=\frac{\dn\,u_{\theta}\pm ik\,\sn\,u_{\theta}\,\cn\,u_{\theta}}{\sqrt{1-k^2\sn^4u_{\theta}}}.
 \eb

 From (\ref{ellbtheta2}) we can now deduce what we really want --- namely, a nice elliptic representation of the matrix elements of $A$, $B$, $C$, $D$ in Lemma~\ref{lemabcd}:
 \begin{lemma}\label{ellipticabcd} Let $\theta,\theta'\in[0,2\pi)$ with $\theta\pm \theta'\neq0,2\pi$. Then
 \begin{align}
 \label{dmatrix}\frac{1}{2\sin\frac{\theta-\theta'}{2}}\left(\sqrt{\frac{b_{\theta}}{b_{\theta'}}}+
 \sqrt{\frac{b_{\theta'}}{b_{\theta}}}\right)\quad\;=&\,
 \frac{\sinh2\ky}{\sqrt{\sinh\gamma_{\theta}\sinh\gamma_{\theta'}}}
 \frac{\dn(u_{\theta}-u_{\theta'})}{\sn(u_{\theta}-u_{\theta'})},\\
 \label{cmatrix}\frac{1}{2\sin\frac{\theta+\theta'}{2}}
 \left(\sqrt{b_{\theta}b_{\theta'}}-\frac{1}{\sqrt{b_{\theta}b_{\theta'}}}\right)=&\,
 \frac{-i\sinh2\kx^*}{\sqrt{\sinh\gamma_{\theta}\sinh\gamma_{\theta'}}}
 \,\cn(u_{\theta}-u_{\theta'}).
 \end{align}
 \end{lemma}
 \pf From (\ref{ellbtheta2}) it follows that e.g.
 \ben
 \frac{1}{2\sin\frac{\theta-\theta'}{2}}\left(\sqrt{\frac{b_{\theta}}{b_{\theta'}}}+
 \sqrt{\frac{b_{\theta'}}{b_{\theta}}}\right)=\frac{\dn\,u_{\theta}\,\dn\,u_{\theta'}+
 k^2\sn\,u_{\theta}\,\cn\,u_{\theta}\,\sn\,u_{\theta'}\,\cn\,u_{\theta'}}{
 \sin\frac{\theta-\theta'}{2}\,\sqrt{\left(1-k^2\sn^4u_{\theta}\right)\left(1-k^2\sn^4u_{\theta'}\right)}}.
 \ebn
 Standard addition formula $\displaystyle\dn(u-u')=\frac{\dn\,u\,\dn\,u'+k^2\sn\,u\,\cn\,u\,\sn\,u'\,\cn\,u'}{
 1-k^2\sn^2u\,\sn^2u'}$ transforms the right side of the last relation into
 \ben
 \frac{1-k^2\sn^2u_{\theta}\sn^2u_{\theta'}}{
 \sqrt{\left(1-k^2\sn^4u_{\theta}\right)\left(1-k^2\sn^4u_{\theta'}\right)}}\,\frac{
 \dn(u_{\theta}-u_{\theta'})}{\sin\frac{\theta-\theta'}{2}}.
 \ebn
 By (\ref{snaux4}), the first factor may be reduced to $\displaystyle
 \frac{\sinh\frac{\gamma_{\theta}+\gamma_{\theta'}}{2}}{\sqrt{\sinh\gamma_{\theta}\sinh\gamma_{\theta'}}}$,
 and the desired identity (\ref{dmatrix}) becomes an immediate consequence of (\ref{snttp}).

 Similarly, applying (\ref{ellbtheta2}) to the l.h.s. of (\ref{cmatrix}) we get
 \ben
 \frac{1}{2\sin\frac{\theta+\theta'}{2}}
 \left(\sqrt{b_{\theta}b_{\theta'}}-\frac{1}{\sqrt{b_{\theta}b_{\theta'}}}\right)=
 ik\frac{\sn\,u_{\theta}\,\cn\,u_{\theta}\,\dn\,u_{\theta'}+\dn\,u_{\theta}\,\sn\,u_{\theta'}\,\cn\,u_{\theta'}}{
 \sin\frac{\theta+\theta'}{2}\,\sqrt{\left(1-k^2\sn^4u_{\theta}\right)\left(1-k^2\sn^4u_{\theta'}\right)}}.
 \ebn
 Now (\ref{cmatrix}) can be obtained from the addition formula
 \ben\displaystyle \cn(u-u')\,\sn(u+u')=
 \frac{\sn\,u\,\cn\,u\,\dn\,u'+\dn\,u\,\sn\,u'\,\cn\,u'}{
 1-k^2\sn^2u\,\sn^2u'}
 \ebn
 in a way analogous to the above.
  \epf

  Finally, let us introduce two diagonal matrices $\Lambda_{\nu}$ ($\nu=a,p$) with elements
  \be\label{lambdaap}
  \left(\Lambda_{\nu}\right)_{\theta,\theta'}=
  \frac{e^{i(l-\frac12)\theta}}{\sqrt{\sinh\gamma_{\theta}}}\,\delta_{\theta,\theta'},\qquad
  \theta,\theta'\in\boldsymbol{\theta}_{\nu},
  \eb
  and two non-diagonal matrices $\Phi$, $\Psi$ defined by
  \begin{align}\label{phipsidef}
  \Phi_{\theta,\theta'}=\frac{\dn(u_{\theta}-u_{\theta'})}{\sn(u_{\theta}-u_{\theta'})},\qquad
  \Psi_{\theta,\theta'}=\cn(u_{\theta}-u_{\theta'}),\qquad
  \theta\in\boldsymbol{\theta}_p,\theta'\in\boldsymbol{\theta}_a.
  \end{align}
  Combining the results of Lemmas~\ref{lemabcd} and~\ref{ellipticabcd}, we arrive at
  \begin{lemma}\label{lemsff} Matrices $B$, $C$ and $D$ defined by (\ref{elemsad})--(\ref{elemsbc})
  satisfy the following identities:
  \begin{align}
  \label{xxx1}D^{-1}=&\,\frac{-iN}{\sinh2\ky}\,\Lambda_a^{-1}\Phi^{-1}\bar{\Lambda}_p^{-1},\\
  \label{xxx2}BD^{-1}=&\,i\,\frac{\sinh2\kx^*}{\sinh2\ky}\,\Lambda_p\Psi\Phi^{-1}\bar{\Lambda}_p^{-1}, \\
  \label{xxx3}D^{-1}C=&\,i\,\frac{\sinh2\kx^*}{\sinh2\ky}\,\Lambda_a^{-1}\Phi^{-1}\Psi\bar{\Lambda}_a,\\
  \label{xxx4}|\mathrm{det}\,D|=&\,\frac{\left(\sinh2\ky\right)^N|\mathrm{det}\,\Phi|}{N^N\sqrt{
  \prod_{\theta\in\boldsymbol{\theta}_p}\sinh\gamma_{\theta}
  \prod_{\theta\in\boldsymbol{\theta}_a}\sinh\gamma_{\theta}}}\,.
  \end{align}
  \end{lemma}
  \pf Check that $\displaystyle B=\bar{C}=-\frac{\sinh2\kx^*}{N}\,\Lambda_p\Psi\Lambda_a$, $\displaystyle D=-\frac{\sinh2\ky}{iN}\,\bar{\Lambda}_p\Phi\Lambda_a$.\epf

  Thus in order to obtain all finite-lattice form factors of Ising spin,
  it is sufficient to compute the inverse $\Phi^{-1}$ and two matrix products
  $\Psi\Phi^{-1}$, $\Phi^{-1}\Psi$. This will be done in the next section
  using that $\Phi$, $\Psi$ defined by (\ref{phipsidef}) are special cases of elliptic Cauchy matrices.

 \section{Elliptic Cauchy matrices and two-particle form factors\label{sec_cauchy}}
 \subsection{Elliptic determinants}
 Consider $2N$ complex variables $x_1,\ldots,x_N$, $y_1,\ldots,y_N$. The
 determinant of the Cauchy matrix $\displaystyle V_{ij}=\frac{1}{x_i-y_j}$ ($i,j=1,\ldots,N$) is a rational
 function of $\{x_i\}$ and $\{y_i\}$, which vanishes whenever $x_i=x_j$ or $y_i=y_j$ for some pair of distinct indices $i,j$.
 It is well-known that there are no other zeros and the determinant is given by
 \ben
 \mathrm{det}\,V=\frac{\prod_{i<j}^N(x_i-x_j)(y_j-y_i)}{\prod^N_{i,j}(x_i-y_j)}.
 \ebn

 There is an elliptic version of this relation due to Frobenius \cite{Frobenius}. Namely, if we denote
 by $\vartheta_{1\ldots 4}(z)$ Jacobi theta functions of nome $q=e^{\pi i \tau}$, then
 for the elliptic Cauchy matrix
 $\displaystyle\tilde{V}_{ij}=\frac{\vartheta_1(x_i-y_j+\alpha)}{\vartheta_1(x_i-y_j)\vartheta_1(\alpha)}$
 depending on an arbitrary parameter $\alpha$ one has the identity
 \begin{align}
 \label{dettheta}
 \mathrm{det}\,\tilde{V}=
 \frac{\vartheta_1\left(\sum^N_i x_i-\sum^N_i y_i+\alpha\right)}{\vartheta_1(\alpha)}
 \frac{\prod\nolimits^N_{i<j}\vartheta_1(x_i-x_j)\vartheta_1(y_j-y_i)}{\prod\nolimits^N_{i,j}\vartheta_1(x_i-y_j)}.
 \end{align}
 \begin{corollary} We have
 \begin{align}
 \label{vmone}\tilde{V}^{-1}_{mn}=&\,-\frac{\vartheta_1\left(\sum^N_{i\neq n}x_i-\sum^N_{i\neq m}y_i+\alpha\right)}{
 \vartheta_1\left(\sum_{i}^Nx_i-\sum^N_{i}y_i+\alpha\right)\vartheta_1\left(x_n-y_m\right)}\times\\
 \nonumber &\,\times\frac{\prod^N_i\vartheta_1\left(x_n-y_i\right)
 \prod^N_i\vartheta_1\left(y_m-x_i\right)}{
 \prod^N_{i\neq n}\vartheta_1\left(x_n-x_i\right)
 \prod^N_{i\neq m}\vartheta_1\left(y_m-y_i\right)}.
 \end{align}
 \end{corollary}
 \pf Use (\ref{dettheta}) to compute the determinant and cofactors and (\ref{vmone}) follows. \epf

 Theta functions are related to the Jacobi elliptic functions by
 \be\label{thetaelliptic}
 \sn\,u=\frac{\vartheta_3}{\vartheta_2}\frac{\vartheta_1(\vartheta_3^{-2}u)}{\vartheta_4(\vartheta_3^{-2}u)},\qquad
 \cn\,u=\frac{\vartheta_4}{\vartheta_2}\frac{\vartheta_2(\vartheta_3^{-2}u)}{\vartheta_4(\vartheta_3^{-2}u)},\qquad
 \dn\,u=\frac{\vartheta_4}{\vartheta_3}\frac{\vartheta_3(\vartheta_3^{-2}u)}{\vartheta_4(\vartheta_3^{-2}u)},
 \eb
 where $\vartheta_i=\vartheta_i(0)$ for $i=2,3,4$. Elliptic modulus and half-periods are given by
 \ben
 k=\frac{\vartheta_2^2}{\vartheta_3^2},\qquad 2K=\pi\vartheta_3^2,\qquad 2iK'=\pi\tau\vartheta_3^2.
 \ebn
 The functions $\vartheta_{2,3,4}(z)$ can be obtained from $\vartheta_1(z)$ by shifting its argument,
 \begin{align}
 \label{theta2}\vartheta_2(z)=&\,\vartheta_1\left(z+\frac{\pi}{2}\right),\\
 \label{theta3}\vartheta_3(z)=&\,e^{-i\left(z-\frac{\pi\tau}{4}\right)}\vartheta_1\left(z+\frac{\pi}{2}-\frac{\pi\tau}{2}\right),\\
 \label{theta4}\vartheta_4(z)=&\,ie^{-i\left(z-\frac{\pi\tau}{4}\right)}\vartheta_1\left(z-\frac{\pi\tau}{2}\right).
 \end{align}
 Therefore one can deduce from (\ref{dettheta}), (\ref{vmone}) the determinants and inverses
 for a number of matrices with entries written in terms of the Jacobi elliptic functions.
 \begin{lemma} For $1\leq i,j\leq N$ with even  $N$ we have
 \be\label{detsn}
 \mathrm{det}\left(\sqrt{k}\;\sn(u_i-u_j)\right)=\left(\prod^N_{i<j}\sqrt{k}\;\sn(u_i-u_j)\right)^2.
 \eb
 \end{lemma}
 \pf Set  $x_i=\vartheta_3^{-2}u_i$, $y_i=\vartheta_3^{-2}u_i+\frac{\pi\tau}{2}$
 and use (\ref{thetaelliptic})--(\ref{theta4}) and the formula $\vartheta_1\left(\frac{\pi\tau}{2}\right)
 =ie^{-\frac{i\pi\tau}{4}}\vartheta_4$
 to write
 \ben
 \sqrt{k}\,\sn(u_i-u_j)=\vartheta_4 e^{i(x_i-y_j)}\frac{\vartheta_1\left(x_i-y_j+\frac{\pi\tau}{2}\right)}{
 \vartheta_1\left(x_i-y_j\right)\vartheta_1\left(\frac{\pi\tau}{2}\right)}.
 \ebn
 The determinant on the left side of (\ref{detsn})
 can therefore be deduced from (\ref{dettheta}) with $\alpha=\frac{\pi\tau}{2}$.
 Since in our case $y_i=x_i+\frac{\pi\tau}{2}$ and in particular
 $\sum^N_ix_i-\sum^N_iy_i=-\frac{N\pi\tau}{2}$, it reduces to
 \ben
 (-1)^{\frac{N}{2}-1}e^{-\frac{iN\pi\tau}{4}}\frac{\vartheta_1\bigl(\frac{(N-1)\pi\tau}{2}\bigr)}{
 \vartheta_1\left(\frac{\pi\tau}{2}\right)}\prod^N_{i<j}\frac{\vartheta_1(x_i-x_j)\vartheta_1(x_j-x_i)}{
 \vartheta_1\left(x_i-x_j-\frac{\pi\tau}{2}\right)\vartheta_1\left(x_j-x_i-\frac{\pi\tau}{2}\right)}.
 \ebn
 To transform the
 theta functions back to $\mathrm{sn}$'s, use again (\ref{thetaelliptic})--(\ref{theta4}) and the quasiperiodicity
  relation $\vartheta_1(z+\ell\pi\tau)=(-1)^{\ell}e^{-i\pi\ell^2\tau-2i\ell z}\vartheta_1(z)$, which holds for $\ell\in\mathbb{Z}$. \epf

  \begin{remark} Taking the square root at both sides of (\ref{detsn}) and fixing the signs via
  the rational limit, one obtains a pfaffian version of this formula:
  \be\label{pfsn}
 \mathrm{Pf}\left(\sqrt{k}\;\sn(u_i-u_j)\right)=\prod^N_{i<j}\sqrt{k}\;\sn(u_i-u_j).
 \eb
 We note that this identity has already appeared in the proof of Theorem~5.0 in \cite{PT},
 where it was used to obtain multiparticle Ising form factors on the infinite
 lattice from the two-particle ones.
  \end{remark}

  \begin{lemma}\label{lemdnsn} Let $\Phi$ denote a matrix with elements
 $\displaystyle\Phi_{ij}=\frac{\dn(u_i-v_j)}{\sn(u_i-v_j)}$ with
 $i,j=1,\ldots,N$. Then
 \begin{align}
 \nonumber\mathrm{det}\,\Phi=&\,\frac{\vartheta_2^N\vartheta_4^N}{\vartheta_3^{N+1}}\,
 e^{-i\left(\sum^N_ix_i-\sum^N_i y_i-\frac{\pi \tau}{4}\right)}\vartheta_1\left(\smsum^N_i x_i-\smsum^N_i y_i+\frac{\pi}{2}-\frac{\pi\tau}{2}\right)\times\\
  \label{sndndet}&\,\times
 \frac{\prod\nolimits^N_{i<j}\vartheta_1(x_i-x_j)\vartheta_1(y_j-y_i)}{
 \prod\nolimits^N_{i,j}\vartheta_1(x_i-y_j)},\\
 \nonumber
 \Phi^{-1}_{mn}=&\,-\frac{\vartheta_3}{\vartheta_2\vartheta_4}\frac{e^{i(x_n-y_m)}\vartheta_1\left(\sum^N_{i\neq n}x_i-\sum^N_{i\neq m}y_i+\frac{\pi}{2}-\frac{\pi\tau}{2}\right)}{
 \vartheta_1\left(\sum^N_{i}x_i-\sum^N_{i}y_i+\frac{\pi}{2}-\frac{\pi\tau}{2}\right) \vartheta_1\left(x_n-y_m\right)}\,\times\\  \label{sm1} &\,\times\frac{\prod^N_i\vartheta_1\left(x_n-y_i\right)
 \prod^N_i\vartheta_1\left(y_m-x_i\right)}{
 \prod^N_{i\neq n}\vartheta_1\left(x_n-x_i\right)
 \prod^N_{i\neq m}\vartheta_1\left(y_m-y_i\right)},
 \end{align}
 where $x_i=\vartheta_3^{-2}u_i$, $y_i=\vartheta_3^{-2}v_i$.
 \end{lemma}
 \pf From (\ref{thetaelliptic}) and the relation
 $\vartheta_1\left(\frac{\pi}{2}-\frac{\pi\tau}{2}\right)=e^{-\frac{i\pi \tau}{4}}\vartheta_3$
 it follows that $\Phi_{ij}$ can be written as
 \ben
 \Phi_{ij}=e^{-i(x_i-y_j)}\frac{\vartheta_2\vartheta_4}{\vartheta_3}\frac{\vartheta_1\left(x_i-y_j+
 \frac{\pi}{2}-\frac{\pi\tau}{2}\right)}{\vartheta_1(x_i-y_j)\,\vartheta_1\left(
 \frac{\pi}{2}-\frac{\pi\tau}{2}\right)}.
 \ebn
 Now to derive (\ref{sndndet}), (\ref{sm1}), it suffices to set
  $\alpha= \frac{\pi}{2}-\frac{\pi\tau}{2}$ in (\ref{dettheta}), (\ref{vmone}). \epf

 \subsection{Theta functional interpolation} Let $z_1,\ldots,z_M$, $f_1,\ldots,f_M$ be $2M$ complex variables.
 The Lagrange polynomial $P(z)$ is the unique polynomial of degree $\leq M-1$, interpolating the set of points $(z_i,f_i)$ (in
 other words, $f_i=P(z_i)$ for $i=1,\ldots,M$). As is well-known,
 \be\label{lagrange}
 P(z)=\sum_{i}^M f_i\prod_{j\neq i}^M\frac{z-z_j}{z_i-z_j}.
 \eb
 Introduce $M$ additional parameters $z_1',\ldots,z_M'$ and the $(M-1)$th degree polynomial
 \ben f(z)=\prod\limits_{i}^M(z-z_i')-
 \prod\limits_{i}^M(z-z_i).
 \ebn
 If we now set $f_j=f(z_j)$ for $j=1,\ldots,M$, then the evaluation of the coefficient of $z^{M-1}$ at both
 sides of (\ref{lagrange}) leads to the identity
 \ben
 \sum\limits_i^M\frac{\prod^M_{j}(z_i-z_j')}{\prod^M_{j\neq i} (z_i-z_j)}=\sum_{i}^Mz_i-\sum_{i}^Mz_i'.
 \ebn

 The last relation has an elliptic analog, which turns out to be very helpful in the
 calculation of the products $BD^{-1}$ and $D^{-1}C$. Namely, if the parameters $\{z_i\}$, $\{z_i'\}$
 satisfy the balancing condition $\sum\limits_{i}^Mz_i-\sum\limits_{i}^Mz_i'=0$, then
 \be\label{thetainterpolation}
  \sum\limits_i^M\frac{\prod_{j}^M\vartheta_1(z_i-z_j')}{\prod^M_{j\neq i} \vartheta_1(z_i-z_j)}=0.
 \eb
 This identity appears in a somewhat disguised form in \cite{WW}, Example~3 on p.~451,
 where it is formulated in terms of the Weierstrass $\sigma$-function. It can be
 proved by induction on $M$, using the Riemann's addition formula in the base case,
 see e.g. Theorem~7 in \cite{Spiridonov}. Another proof, based on the Frobenius determinant
 formula, is given in the Appendix of \cite{Nijhoff2}. The identities
 of type (\ref{thetainterpolation}) arise in \cite{Nijhoff1,Nijhoff2}
 in the analysis of a discrete time version of the Ruijsenaars-Schneider
 model.
   \begin{lemma}
 Let $\Psi$ denote a matrix with elements
 $\displaystyle\Psi_{ij}=\cn(u_i-v_j)$ ($i,j=1,\ldots ,N$) and $\Phi$, $\{x_i\}$, $\{y_i\}$ be as in Lemma~\ref{lemdnsn}. Then we have
 \begin{align}
 \nonumber\left(\Psi\Phi^{-1}\right)_{ln}=&\, ie^{i\left(x_n-x_l-\frac{\pi\tau}{2}\right)}\frac{\vartheta_3^2}{\vartheta_2^2}
 \frac{\vartheta_1\left(x_l-x_n+\sum^N_{i}x_i-\sum^N_{i}y_i+\frac{\pi}{2}\right)}{
 \vartheta_1\left(\sum^N_{i}x_i-\sum^N_{i}y_i+\frac{\pi}{2}-\frac{\pi\tau}{2}\right)}\times\\
 \label{psiphi}&\,
 \times\prod\limits^N_{i}\frac{\vartheta_1\left(x_n-y_i\right)}{\vartheta_1\left(x_l-y_i+\frac{\pi\tau}{2}\right)}
 \prod\limits^N_{i\neq n}\frac{\vartheta_1\left(x_l-x_i+\frac{\pi\tau}{2}\right)}{\vartheta_1\left(x_n-x_i\right)},\\
 \nonumber\left(\Phi^{-1}\Psi\right)_{ml}=&\,ie^{i\left(y_l-y_m-\frac{\pi\tau}{2}\right)}\frac{\vartheta_3^2}{\vartheta_2^2}
 \frac{\vartheta_1\left(y_m-y_l+\sum^N_{i}x_i-\sum^N_{i}y_i+\frac{\pi}{2}\right)}{
 \vartheta_1\left(\sum^N_{i}x_i-\sum^N_{i}y_i+\frac{\pi}{2}-\frac{\pi\tau}{2}\right)}\times\\
 \label{phipsi}&\,
 \times\prod\limits^N_{i}\frac{\vartheta_1\left(y_m-x_i\right)}{\vartheta_1\left(y_l-x_i-\frac{\pi\tau}{2}\right)}
 \prod\limits^N_{i\neq m}\frac{\vartheta_1\left(y_l-y_i-\frac{\pi\tau}{2}\right)}{\vartheta_1\left(y_m-y_i\right)}.
 \end{align}
 \end{lemma}
 \pf We will prove only the first identity, since for the second the argument is
 completely analogous. By (\ref{thetaelliptic}), one can write
 \ben
 \Psi_{lm}=-ie^{i\left(y_m-x_l-\frac{\pi\tau}{4}\right)}\frac{\vartheta_4}{\vartheta_2}
 \frac{\vartheta_1\left(y_m-x_l+\frac{\pi}{2}\right)}{\vartheta_1\left(y_m-x_l-\frac{\pi\tau}{2}\right)}.
 \ebn
 Lemma~\ref{lemdnsn} then implies that the l.h.s. of (\ref{psiphi}) is given by
 \begin{align}
 \nonumber
 \left(\Psi\Phi^{-1}\right)_{ln}=&\,\frac{\vartheta_3}{\vartheta_2^2}\frac{ie^{i\left(x_n-x_l-\frac{\pi\tau}{4}\right)}}{
 \vartheta_1\left(\sum^N_{i}x_i-\sum^N_{i}y_i+\frac{\pi}{2}-\frac{\pi\tau}{2}\right)}
 \frac{\prod^N_i\vartheta_1(x_n-y_i)}{\prod^N_{i\neq n}\vartheta_1(x_n-x_i)}\,\times\\
 \label{auxpsiphi}&\,\times\,\sum\limits_m^N\biggl\{
 \vartheta_1\left(\smsum^N_{i\neq n}x_i-\smsum^N_{i\neq m}y_i-\frac{\pi}{2}-\frac{\pi\tau}{2}\right)
 \biggr.\times\\
 \nonumber&\,\times\biggl.\frac{\vartheta_1\left(y_m-x_l+\frac{\pi}{2}\right)}{\vartheta_1\left(y_m-x_l-\frac{\pi\tau}{2}\right)}
 \frac{\prod^N_{i\neq n}\vartheta_1(y_m-x_i)}{\prod^N_{i\neq m}\vartheta_1(y_m-y_i)}\biggr\}.
 \end{align}
 If we denote
 \begin{description}
 \item $z_i=y_i$ for $i=1,\ldots,N$, $z_{N+1}=x_l+\frac{\pi\tau}{2}$,
 \item $z'_i=x_i$ for $i=1,\ldots,n-1$ and for $i=n+1,\ldots,N$,
 \item $z'_n=x_n-\sum\nolimits^N_ix_i+\sum\nolimits^N_iy_i+\frac{\pi}{2}+\frac{\pi\tau}{2}$ and $z'_{N+1}=x_l-\frac{\pi}{2}$,
 \end{description}
 then the sum in  (\ref{auxpsiphi}) can be written as
 $\displaystyle
 \sum\limits_i^{N}\frac{\prod^{N+1}_{j}\vartheta_1(z_i-z_j')}{\prod^{N+1}_{j\neq i} \vartheta_1(z_i-z_j)}$. This almost coincides
 with the sum in the l.h.s. of (\ref{thetainterpolation}) with $M=N+1$.
 The missing $(N+1)$th term is given by
 \begin{align*}
 \frac{\prod_i^{N+1}\vartheta_1(z_{N+1}-z'_j)}{\prod_i^N\vartheta_1(z_{N+1}-z_j)}=&\,
 \vartheta_1\left(\frac{\pi}{2}+\frac{\pi\tau}{2}\right)\vartheta_1\left(x_l-x_n
 +\text{\footnotesize$\sum\limits_i^N$}x_i-\text{\footnotesize$\sum\limits_i^N$}y_i-
 \frac{\pi}{2}\right)\times
 \\
 &\,\times\frac{\prod_{i\neq n}^N\vartheta_1\left(
 x_l-x_i+\frac{\pi\tau}{2}\right)}{\prod_{i}^N\vartheta_1\left(x_l-y_i+\frac{\pi\tau}{2}\right)}.
 \end{align*}
 Since the balancing condition is satisfied, our sum is equal to minus this missing term. Together
 with the first line in (\ref{auxpsiphi}), this leads to (\ref{psiphi}).
 \epf
 \subsection{Trigonometric reduction}
 The matrices $\Phi$ and $\Psi$, corresponding to the Ising case, are indexed by
 two sets of quasimomenta. Explicitly, $u_j=u_{\theta}$ with
 $\theta=\frac{2\pi(j-1)}{N}\in\boldsymbol{\theta}_{p}$, and
 $v_j= u_{\theta'}$ with
 $\theta'=\frac{2\pi}{N}\left(j-\frac12\right)\in\boldsymbol{\theta}_a$.
 Since for $\theta\in(0,2\pi)$ each set contains both
 $\theta$ and $2\pi-\theta$, there are additional constraints on $\{x_j\}$, $\{y_j\}$. Their form slightly
 differs for odd and even values of $N$:
 \begin{itemize}
 \item For odd $N$, the value $\theta=0$ belongs to the periodic set of momenta,
 but $\theta=\pi$ is in the
 antiperiodic one. Then for all $j=2,\ldots,\frac{N+1}{2}$ and $k=1,\ldots,\frac{N-1}{2}$ we have
    \be\label{constraintodd}
     x_j+x_{N+2-j}=y_k+y_{N+1-k}=x_1+\frac{\pi}{2}=y_{(N+1)/2}=0.
     \eb
 \item For even $N$, both exceptional values $\theta=0,\pi$ belong to the periodic set. Thus
  for all $j=2,\ldots,\frac{N}{2}$ and $k=1,\ldots,\frac{N}{2}$ one has
     \be\label{constrainteven}
     x_j+x_{N+2-j}=y_k+y_{N+1-k}=x_1+\frac{\pi}{2}=x_{N/2+1}=0.
     \eb
 \end{itemize}

 In particular, in both cases we have
 \be\label{sumconstraint}
 \sum_i^Nx_i-\sum_i^Ny_i=-\frac{\pi}{2}.
 \eb
 This allows to rewrite
 (\ref{sndndet})--(\ref{sm1}) and (\ref{psiphi})--(\ref{phipsi}) as
 \be\label{detandinversephi}
 \left(\mathrm{det}\,\Phi\right)^2=\frac{\vartheta_3^{-2}\vartheta_4^{2}}{
 \prod_{i}^Nf_ig_i},\qquad \Phi^{-1}_{mn}=\frac{
 f_ng_m}{\sn(u_n-v_m)},
 \eb
   \begin{align}\label{psiphi2}
 \left(\Psi\Phi^{-1}\right)_{ln}=&\,
 f_n h\bigl(x_l+\frac{\pi\tau}{2}\bigr)\,\sn(u_l-u_n),
 \\
 \label{phipsi2}\left(\Phi^{-1}\Psi\right)_{ml}=&\,
 -\frac{g_m}{h\left(y_l-\frac{\pi\tau}{2}\right)}\,\sn(v_m-v_l),
 \end{align}
 where $\displaystyle
 h(z)=\prod_{i}^N\frac{\vartheta_1(z-x_i)}{\vartheta_1(z-y_i)}$ and $\{f_n\}$, $\{g_m\}$  are defined by
 \ben
 f_n=\frac{i\vartheta_3}{\vartheta_2\vartheta_4}\frac{\;\prod_{i}^N\vartheta_1(x_n-y_i)}{\prod_{i\neq n}^N\vartheta_1(x_n-x_i)},
 \qquad
 g_m=\frac{i\vartheta_3}{\vartheta_2\vartheta_4}\frac{\;\prod_{i}^N\vartheta_1(y_m-x_i)}{\prod_{i\neq m}^N\vartheta_1(y_m-y_i)}.
 \ebn

 Notice that (\ref{sumconstraint}), combined with the quasiperiodicity of $\vartheta_1(z)$,
 also implies that
 $h(z+\pi\tau)=-h(z)$. Further, it is easy to check using (\ref{thetaelliptic}), (\ref{theta4}) that
 the quantities $\chi_n=-f_nh\bigl(x_n-\frac{\pi\tau}{2}\bigr)$ and
 $\kappa_m=g_m/h\left(y_m-\frac{\pi\tau}{2}\right)$ are given by
 \begin{align}
 \label{fnhgmh}
 \chi_n=\frac{\,\prod_{i}^N\sn(u_n-v_i)}{
 \prod_{i\neq n}^N\sn(u_n-u_i)},\qquad
 \kappa_m=\frac{\;\prod_{i}^N\sn(v_m-u_i)}{
 \prod_{i\neq m}^N\sn(v_m-v_i)}.
 \end{align}
 Thus, to write (\ref{detandinversephi})--(\ref{phipsi2}) in terms of sn's, it suffices
 to find a suitable representation for the function $\mu(z,z')=h(z)/h(z')$. For that,
 one can use the constraints (\ref{constraintodd})--(\ref{constrainteven}).

 Suppose e.g. that
 $N$ is even, then
 \be\label{muzzpr}
 \mu(z,z')=\frac{\vartheta_1(z)\vartheta_2(z)}{\vartheta_1(z')\vartheta_2(z')}
 \prod_{i=2}^{N/2}\frac{\vartheta_1(z-x_i)\vartheta_1(z+x_i)}{\vartheta_1(z'-x_i)\vartheta_1(z'+x_i)}
  \prod_{i=1}^{N/2}\frac{\vartheta_1(z'-y_i)\vartheta_1(z'+y_i)}{\vartheta_1(z-y_i)\vartheta_1(z+y_i)},
 \eb
 where we have put together the theta functions with arguments coming in pairs according
 to (\ref{constrainteven}). In each product of two theta functions, use the well-known addition formula,
 written in the form
 \begin{align}
 \nonumber\vartheta_1(s+t)\vartheta_1(s-t)\vartheta_4^2=&\,\vartheta_1^2(s)\vartheta_4^2(t)-
 \vartheta_4^2(s)\vartheta_1^2(t)=\\
 \label{addtheta1}=&\,\vartheta_1(s)\vartheta_4(s)\vartheta_1(t)\vartheta_4(t)
 \frac{\sn^2\vartheta_3^2s-\sn^2\vartheta_3^2t}{\sn\,\vartheta_3^2s\;\sn\,\vartheta_3^2t}.
 \end{align}
 After substitution into (\ref{muzzpr}), almost all of the theta functions and denominators appearing
 in (\ref{addtheta1}) cancel each other, so that $\mu(z,z')$ reduces to
 \be\label{musn}
 \mu(z,z')=\frac{\cn\,\vartheta_3^2z\left(1+\sn\,\vartheta_3^2z'\right)}{
 \cn\,\vartheta_3^2z'\left(1+\sn\,\vartheta_3^2z\right)}\prod_{i=1}^N\frac{
 \left(\sn\,\vartheta_3^2z-\sn\,u_i\right)\left(\sn\,\vartheta_3^2z'-\sn\,v_i\right)}{
 \left(\sn\,\vartheta_3^2z-\sn\,v_i\right)\left(\sn\,\vartheta_3^2z'-\sn\,u_i\right)}.
 \eb
 Similar manipulations for odd $N$ lead to the same formula.

 If we now denote $\lambda(u,v)=\mu\left(\vartheta_3^{-2}u-\frac{\pi\tau}{2},
 \vartheta_3^{-2}v-\frac{\pi\tau}{2}\right)$, then by (\ref{musn})
 \be\label{lambdasn}
 \lambda(u,v)=\frac{\dn\,u\left(1+k\,\sn\,v\right)}{
 \dn\,v\left(1+k\,\sn\,u\right)}\prod_{i=1}^N\frac{
 \left(1-k\,\sn\,u_i\,\sn\,u\right)\left(1-k\,\sn\,v_i\,\sn\,v\right)}{
 \left(1-k\,\sn\,v_i\,\sn\,u\right)\left(1-k\,\sn\,u_i\,\sn\,v\right)}.
 \eb
  Relations (\ref{fnhgmh}) and (\ref{lambdasn}) then allow to rewrite (\ref{detandinversephi})--(\ref{phipsi2})
 in terms of the Jacobi elliptic functions:
 \be\label{detphisn}
 \left(\mathrm{det}\,\Phi\right)^2=\frac{(-1)^N\sqrt{1-k^2}}{
 \prod_{i}^N\kappa_i\chi_i\,\lambda(v_i,u_i)},\qquad\Phi^{-1}_{mn}=\frac{ \kappa_m\chi_n\lambda(v_m,u_n)}{\sn(v_m-u_n)},
 \eb
 \begin{align}
 \label{psiphi3}\left(\Psi\Phi^{-1}\right)_{ln}=&\,
 \chi_n\lambda(u_l,u_n)\,\sn(u_l-u_n),\\
 \label{phipsi3}\left(\Phi^{-1}\Psi\right)_{ml}=&\,\kappa_m\lambda(v_m,v_l)\,\sn(v_l-v_m).
 \end{align}

 So far we have used only a reflection symmetry of the periodic and antiperiodic spectrum of momenta.
 That these momenta take rather special values (recall that $e^{i\theta}$ is an $N$th root of $\pm1$)
 leads to further
 simplifications. It will be convenient to switch the notation and write all matrix indices
 in terms of momenta, as suggested in the
 beginning of this subsection. Thus, e.g.
 \be\label{chikappatheta}
 \chi_{\theta}=\frac{\prod_{\theta'\in\boldsymbol{\theta}_a}\sn(u_{\theta}-u_{\theta'})}{
 \prod_{\theta'\in\boldsymbol{\theta}_p,\theta'\neq\theta}\sn(u_{\theta}-u_{\theta'})},
 \qquad
 \kappa_{\theta}=\frac{\prod_{\theta'\in\boldsymbol{\theta}_p}\sn(u_{\theta}-u_{\theta'})}{
 \prod_{\theta'\in\boldsymbol{\theta}_a,\theta'\neq\theta}\sn(u_{\theta}-u_{\theta'})},
 \eb
 where in the first formula $\theta\in\boldsymbol{\theta}_p$ and in the second one $\theta\in\boldsymbol{\theta}_a$. Likewise, for $\theta,\theta'\in [0,2\pi)$ denote
 \be\label{lambdaththpr}
 \lambda_{\theta,\theta'}=\lambda(u_{\theta},u_{\theta'})=
 \frac{\dn\,u_{\theta}\,}{
 \dn\,u_{\theta'}}
 \prod\limits_{\substack{\theta''\in\boldsymbol{\theta}_p\\ \theta''\neq0\;\,}}\frac{1-k\,\sn\,u_{\theta''}\,\sn\,u_{\theta\;}}{
 1-k\,\sn\,u_{\theta''}\,\sn\,u_{\theta'}}
 \prod\limits_{\theta''\in\boldsymbol{\theta}_a}\frac{1-k\,\sn\,u_{\theta''}\,\sn\,u_{\theta'}}{
 1-k\,\sn\,u_{\theta''}\,\sn\,u_{\theta\;}}.
 \eb

 Our aim is to express $\chi_{\theta}$, $\kappa_{\theta}$ and $\lambda_{\theta,\theta'}$
 in the initial trigonometric pa\-ra\-me\-trization (i.e. in terms of
 $\theta$ and $\gamma_{\theta}$).
 Introduce the function
 \be\label{nutheta}
 \nu_{\theta}=\ln\frac{\prod_{\theta'\in\boldsymbol{\theta}_a}
 \sinh\frac{\gamma_{\theta}+\gamma_{\theta'}}{2}}{\prod_{\theta'\in\boldsymbol{\theta}_p}
 \sinh\frac{\gamma_{\theta}+\gamma_{\theta'}}{2}},\qquad \theta\in [0,2\pi).
 \eb
 Use (\ref{sndn}) to rewrite the first factor in  (\ref{lambdaththpr}) and
 transform the remaining two products with the help of (\ref{snaux4}).
 The calculation is slightly different for odd and even $N$, but the final result is given
 by the same simple formula:
 \be\label{lambdattp2}
 \lambda_{\theta,\theta'}=e^{\left(\nu_{\theta'}-\nu_{\theta}\right)/2}.
 \eb

 Similarly, the identity (\ref{snttp}) allows to write (\ref{chikappatheta}) as
 \begin{align}
 \label{chitheta2}\chi_{\theta}=&\,e^{-\nu_{\theta}}\frac{\sinh2\ky}{\sinh\gamma_{\theta}}
 \frac{\prod_{\theta'\in\boldsymbol{\theta}_a}\sin\frac{\theta-\theta'}{2}}{
 \prod_{\theta'\in\boldsymbol{\theta}_p,\theta'\neq\theta}\sin\frac{\theta-\theta'}{2}}
 \quad\text{for }\theta\in\boldsymbol{\theta}_p,\\
 \label{kappatheta2}\kappa_{\theta}=&\,e^{\,\nu_{\theta}}\;\,\frac{\sinh2\ky}{\sinh\gamma_{\theta}}
 \frac{\prod_{\theta'\in\boldsymbol{\theta}_p}\sin\frac{\theta-\theta'}{2}}{
 \prod_{\theta'\in\boldsymbol{\theta}_a,\theta'\neq\theta}\sin\frac{\theta-\theta'}{2}}
 \quad\text{for }\theta\in\boldsymbol{\theta}_a.
 \end{align}
 The products in the r.h.s. can be easily computed explicitly. One has
 \begin{align}
 \label{prod1}2^{N-1}\prod_{\theta'\in\boldsymbol{\theta}_p}\sin\text{\small{}$\frac{\theta-\theta'}{2}$}=
 &\,(-1)^{N-1}\sin\text{\small{}$\frac{N\theta}{2}$},
 \\
 \label{prod2}2^{N-1}\prod_{\theta'\in\boldsymbol{\theta}_a}
 \sin\text{\small{}$\frac{\theta-\theta'}{2}$}=&\;\;\;
 (-1)^{N}\;\cos \text{\small{}$\frac{N\theta}{2}$}, \\
 \label{prod3}2^{N-1}\prod_{\theta'\in\boldsymbol{\theta}_p,\theta\neq\theta'}
 \sin\text{\small{}$\frac{\theta-\theta'}{2}$}=&\,
 (-1)^{N-1}N\cos \text{\small{}$\frac{N\theta}{2}$}\quad\text{for } \theta\in\boldsymbol{\theta}_p,\\
 \label{prod4} 2^{N-1}\prod_{\theta'\in\boldsymbol{\theta}_a,\theta\neq\theta'}
 \sin\text{\small{}$\frac{\theta-\theta'}{2}$}=&\,
 (-1)^{N-1}N\sin \text{\small{}$\frac{N\theta}{2}$}\quad\text{for } \theta\in\boldsymbol{\theta}_a.
 \end{align}
 The first two relations are valid for any $\theta\in\Cb$ and follow from the obvious
 product formulas $\prod_{\theta'\in\boldsymbol{\theta}_p}(z-e^{i\theta'})=z^N-1$ and
 $\prod_{\theta'\in\boldsymbol{\theta}_a}(z-e^{i\theta'})=z^N+1$. The third and fourth
 formula may be deduced from (\ref{prod1})--(\ref{prod2}) by taking appropriate limits.

 Substituting (\ref{prod1})--(\ref{prod4}) into (\ref{chitheta2})--(\ref{kappatheta2}), and
 combining the result
 with (\ref{snttp}) and (\ref{lambdattp2}), we can rewrite (\ref{detphisn})--(\ref{phipsi3}) as follows:
 \begin{align}
 \label{detphifinal}
 \left(\mathrm{det}\,\Phi\right)^2=&\,\frac{N^{2N}\sqrt{1-k^2}}{\left(\sinh2\ky\right)^{2N}}
 \prod_{\theta\in\boldsymbol{\theta}_p}e^{\nu_{\theta}/2}\sinh\gamma_{\theta}\prod_{\theta\in\boldsymbol{\theta}_a}
 e^{-\nu_{\theta}/2}\sinh\gamma_{\theta},\\
 \label{phimonefinal}\Phi^{-1}_{\theta,\theta'}=&\,\quad-\frac{\sinh2\ky
 e^{\left(\nu_{\theta}-\nu_{\theta'}\right)/2}}{N^2\sinh\gamma_{\theta}\sinh\gamma_{\theta'}}
 \frac{\sinh\frac{\gamma_{\theta}+\gamma_{\theta'}}{2}}{\sin\frac{\theta-\theta'}{2}},\qquad
 \theta\in\boldsymbol{\theta}_a,\theta'\in\boldsymbol{\theta}_p,\\
 \label{psiphifinal}\left(\Psi\Phi^{-1}\right)_{\theta,\theta'}=&\,-\frac{\sinh^22\ky
 e^{-\left(\nu_{\theta}+\nu_{\theta'}\right)/2}}{N\sinh\gamma_{\theta'}}
 \frac{\sin\frac{\theta-\theta'}{2}}{\sinh\frac{\gamma_{\theta}+\gamma_{\theta'}}{2}},\qquad
 \theta,\theta'\in\boldsymbol{\theta}_p,\\
 \label{phipsifinal}\left(\Phi^{-1}\Psi\right)_{\theta,\theta'}=&\,\;\;-\frac{\sinh^22\ky
 e^{\left(\nu_{\theta}+\nu_{\theta'}\right)/2}}{N\sinh\gamma_{\theta}}
 \frac{\sin\frac{\theta-\theta'}{2}}{\sinh\frac{\gamma_{\theta}+\gamma_{\theta'}}{2}},\qquad
 \theta,\theta'\in\boldsymbol{\theta}_a.
 \end{align}

 Now from (\ref{detphifinal})--(\ref{phipsifinal}) and Lemma~\ref{lemsff}
 one easily derives the main result of this paper:
 \begin{theorem}\label{thm2pff}
  Let $B$, $C$ and $D$ be the matrices defined by (\ref{elemsad})--(\ref{elemsbc}). Then
 \begin{align}
 \label{dmone}D^{-1}_{\theta,\theta'}=&\,\quad ie^{-i(l-\frac12)(\theta-\theta')}\frac{
 e^{(\nu_{\theta}-\nu_{\theta'})/2}}{N\sqrt{\sinh\gamma_{\theta}
 \sinh\gamma_{\theta'}}} \frac{\sinh\frac{\gamma_{\theta}+\gamma_{\theta'}}{2}}{\sin\frac{\theta-\theta'}{2}},\\
 \label{bdmonefinal}\left(BD^{-1}\right)_{\theta,\theta'}=&\,\,\,
 -ie^{i(l-\frac12)(\theta+\theta')}\,\frac{\sinh2\ky}{\sinh2\kx}
 \frac{
 e^{-(\nu_{\theta}+\nu_{\theta'})/2}}{N\sqrt{\sinh\gamma_{\theta}
 \sinh\gamma_{\theta'}}} \frac{\sin\frac{\theta-\theta'}{2}}{\sinh\frac{\gamma_{\theta}+\gamma_{\theta'}}{2}},
 \\
 \label{dmonecfinal}\left(D^{-1}C\right)_{\theta,\theta'}=&\,
 -ie^{-i(l-\frac12)(\theta+\theta')}\frac{\sinh2\ky}{\sinh2\kx}
 \frac{
 e^{(\nu_{\theta}+\nu_{\theta'})/2}}{N\sqrt{\sinh\gamma_{\theta}
 \sinh\gamma_{\theta'}}} \frac{\sin\frac{\theta-\theta'}{2}}{\sinh\frac{\gamma_{\theta}+\gamma_{\theta'}}{2}},
 \\
 \label{detdfinal}|\mathrm{det}\,D|=&\;\Bigl[\left(1-k^2\right)\prod_{\theta\in\boldsymbol{\theta}_p}e^{\nu_{\theta}}
 \!\prod_{\theta\in\boldsymbol{\theta}_a}e^{-\nu_{\theta}}\Bigr]^{\frac14},
 \end{align}
 where $\theta\in\boldsymbol{\theta}_a$, $\theta'\in\boldsymbol{\theta}_p$ in (\ref{dmone}),
 $\theta,\theta'\in\boldsymbol{\theta}_p$ in (\ref{bdmonefinal}),  $\theta,\theta'\in\boldsymbol{\theta}_a$
 in (\ref{dmonecfinal}) and the function $\nu_{\theta}$ is given by (\ref{nutheta}).
 \end{theorem}

  By (\ref{vevsigma}) and (\ref{2pffmatr1})--(\ref{2pffmatr3}), Theorem~\ref{thm2pff} gives the vacuum expectation
 value $_a\langle vac|s_l|vac\rangle_p$ and two-particle form factors of Ising spin. Multiparticle
 form factors are determined by Lemma~\ref{mpff}.

 \section{Multiparticle form factors\label{gfsection}}
 We now compute general matrix elements $\mathcal{F}_{m,n}^{(l)}(\boldsymbol{\theta},\boldsymbol{\theta}')$
 using a trick learnt from \cite{Hystad}. There it was shown that, roughly speaking,
 if the conjectures of \cite{BL1,BL2} are true for two-particle form factors,
 then they also hold for multiparticle ones. Theorem~\ref{thm2pff} implies that the last statement
 is no longer conditional.

 The argument of \cite{Hystad} may be sketched as follows. Introduce an $(m+n)\times(m+n)$
 diagonal matrix $\Omega$ with non-zero elements defined by
 \ben
 \Omega_{jj}=\begin{cases}\displaystyle-\frac{\exp\left\{-i(l-\frac12)\theta_j+
 \nu_{\theta_j}/2\right\}}{\sqrt{N\sinh\gamma_{\theta_j}}} & \text{for }j=1,\ldots,m,\vspace{0.2cm}\\
 \displaystyle \frac{\exp\bigl\{i(l-\frac12)\theta'_{j-m}-
 \nu_{\theta'_{j-m}}/2\bigr\}}{\sqrt{N\sinh\gamma_{\theta'_{j-m}}}} & \text{for }j=m+1,\ldots,m+n.\end{cases}
 \ebn
 Taking into account Theorem~\ref{thm2pff}, the skew-symmetric matrix $R$ in Lem\-ma~\ref{mpff} can
 be written as
 \ben
 R=-i\rho\, \Omega\tilde{R}\Omega,\qquad
 \tilde{R}=\left(
 \begin{array}{cc}
 \tilde{R}_{\boldsymbol{\theta}\times\boldsymbol{\theta}} &
 \tilde{R}_{\boldsymbol{\theta}\times\boldsymbol{\theta}'} \\
 \tilde{R}_{\boldsymbol{\theta}'\times\boldsymbol{\theta}} &
 \tilde{R}_{\boldsymbol{\theta}'\times\boldsymbol{\theta}'}
 \end{array}\right)
 \ebn
 where $\displaystyle\rho=\sqrt{\frac{\sinh2\ky}{\sinh2\kx}}$ and matrix elements of the blocks of $\tilde{R}$ are given by
  \begin{align}
 \label{r11v2}\left(\tilde{R}_{\boldsymbol{\theta}\times\boldsymbol{\theta}}\right)_{jk}=
 \frac{\rho\sin\frac{\theta_j-\theta_k}{2}}{\sinh\frac{\gamma_{\theta_j}+\gamma_{\theta_k}}{2}},
 &\qquad j,k=1,\ldots,m,\\
 \label{r12v2} \left(\tilde{R}_{\boldsymbol{\theta}\times\boldsymbol{\theta}'}\right)_{jk}=
 -\left(\tilde{R}_{\boldsymbol{\theta}'\times\boldsymbol{\theta}}\right)_{kj}=
 \frac{\sinh\frac{\gamma_{\theta_j}+\gamma_{\theta'_k}}{2}}{
 \rho\sin\frac{\theta_j-\theta'_k}{2}},
 &\qquad j=1,\ldots,m,\quad k=1,\ldots,n,\\
 \label{r22v2}\left(\tilde{R}_{\boldsymbol{\theta}'\times\boldsymbol{\theta}'}\right)_{jk}=
 \frac{\rho\sin\frac{\theta'_j-\theta'_k}{2}}{\sinh\frac{\gamma_{\theta'_j}+\gamma_{\theta'_k}}{2}},
 &\qquad j,k=1,\ldots,n.
 \end{align}

 Let us recall the identity (\ref{snttp}) from Section~\ref{elpar}, which
 we rewrite in the form
 \be\label{sntthv2}
 \sqrt{k}\,\sn(u_{\theta}-u_{\theta'})=
 \bigl[\sqrt{k}\,\sn(u_{\theta}-u_{\theta'}\pm iK')\bigr]^{-1}=
 \frac{\rho\sin\frac{\theta-\theta'}{2}}{\sinh\frac{\gamma_{\theta}+\gamma_{\theta'}}{2}},
 \qquad \theta,\theta'\in[0,2\pi).
 \eb
 Introducing $(m+n)$ variables
 \be
 \label{util}
 \tilde{u}_j=\begin{cases}u_{\theta_j}+iK' & \text{for }j=1,\ldots,m,\\
 u_{\theta'_{j-m}} & \text{for }j=m+1,\ldots,m+n,\end{cases}
 \eb
 matrix elements of $\tilde{R}$ can be written as $\tilde{R}_{ij}=\sqrt{k}\,\sn(\tilde{u}_i-\tilde{u}_j)$
 with $i,j=1,\ldots, m+n$. The pfaffian (\ref{ffpfaffian}) can then be easily computed using
 the elliptic pfaffian identity (\ref{pfsn}):
 \ben
 \mathrm{Pf}\,R=(-i\rho)^{\frac{m+n}{2}}\cdot\mathrm{det}\,\Omega\,\cdot\mathrm{Pf}\,\tilde{R}=
 (-i\rho)^{\frac{m+n}{2}}\prod_{i=1}^{m+n}\Omega_{ii}\prod_{ i<j}^{m+n}
 \sqrt{k}\,\sn(\tilde{u}_i-\tilde{u}_j).
 \ebn
 Transforming sn's back to trigonometric functions by (\ref{sntthv2})--(\ref{util}) and
 taking into account (\ref{ffpfaffian}) and (\ref{detdfinal}), we finally obtain the desired
 general formula for finite-lattice form factors of Ising spin:
 \begin{theorem} Spin matrix elements $\mathcal{F}_{m,n}^{(l)}(\boldsymbol{\theta},\boldsymbol{\theta}')$,
 defined by (\ref{spinff}), for even $m+n$ can be explicitly written as
 \begin{align}
 \nonumber\mathcal{F}_{m,n}^{(l)}(\boldsymbol{\theta},\boldsymbol{\theta}')=i^{2mn-\frac{ m+n}{2}}
 \sqrt{\xi\xi_T}\times\\
  \label{finalff}\times\left(\frac{\sinh2\ky}{\sinh2\kx}\right)^{\frac{(m-n)^2}{4}}\prod_{j=1}^m
 \frac{e^{-i(l-\frac12)\theta_j+\nu_{\theta_j}/2}}{\sqrt{N\sinh\gamma_{\theta_j}}}
 \prod_{j=1}^n
 \frac{e^{i(l-\frac12)\theta'_j-\nu_{\theta'_j}/2}}{\sqrt{N\sinh\gamma_{\theta'_j}}}\;\times\\
 \nonumber\times\prod_{1\leq i<j\leq m}\frac{\sin\frac{\theta_i-\theta_j}{2}}{\sinh\frac{\gamma_{\theta_i}+\gamma_{\theta_j}}{2}}
  \prod_{1\leq i<j\leq n}\frac{\sin\frac{\theta'_i-\theta'_j}{2}}{\sinh\frac{\gamma_{\theta'_i}+\gamma_{\theta'_j}}{2}}
  \prod_{\substack{1\leq i\leq m \\1\leq j\leq n}}\frac{\sinh\frac{\gamma_{\theta_i}+\gamma_{\theta'_j}}{2}
  }{\sin\frac{\theta_i-\theta'_j}{2}},
 \end{align}
 where $\xi=\left|1-(\sinh2\kx\sinh2\ky)^{-2}\right|^{\frac14}$, the function $\nu_{\theta}$ is defined by (\ref{nutheta}) and
 \be\label{xit}
 \xi_T=\prod_{\theta\in\boldsymbol{\theta}_p} e^{\nu_{\theta}/4}
 \prod_{\theta\in\boldsymbol{\theta}_a} e^{-\nu_{\theta}/4}=
 \left[\frac{\prod_{\theta\in\boldsymbol{\theta}_p}\prod_{\theta'\in\boldsymbol{\theta}_a}
 \sinh^2\frac{\gamma_{\theta}+\gamma_{\theta'}}{2}}{\prod_{\theta,\theta'\in\boldsymbol{\theta}_p}
 \sinh\frac{\gamma_{\theta}+\gamma_{\theta'}}{2}\prod_{\theta,\theta'\in\boldsymbol{\theta}_a}
 \sinh\frac{\gamma_{\theta}+\gamma_{\theta'}}{2}}\right]^{\frac14}.
 \eb
 \end{theorem}

 This result coincides with the conjectures of \cite{BL1,BL2}. It clearly satisfies
 translation invariance constraint (\ref{trinv}). In the thermodynamic limit
  $\xi_T\rightarrow1$, $\nu_{\theta}\rightarrow0$, the spectra of quasimomenta become continuous
 and (\ref{finalff}) reproduces infinite-lattice form factors found in \cite{PT,smj}. In particular,
 considering the $N\rightarrow \infty$ limit of (\ref{detdfinal}), it is straightforward to
  recover Yang's formula $\langle\sigma\rangle=\sqrt{\xi}$
 for the spontaneous magnetization \cite{Yang}.

 \begin{remark} We finally comment on the paramagnetic region of parameters ($\kx^*>\ky$).
 The above results
 remain valid if one performs analytic continuation in e.g. $\kx^*$. Nontrivial dependence of (\ref{finalff})
 on $\kx^*$ is hidden in the functions $\gamma_{\theta}$ with $\theta\in
 \boldsymbol{\theta}_{a,p}$. The result of analytic continuation
 of almost all such $\gamma_{\theta}$ from the region $\kx^*<\ky$ to  $\kx^*>\ky$ coincides with the
 definition (\ref{scurve1}). The only exception is $\gamma_0$, which is continued to $-\gamma_0$.
 Then one finds that under continuation
 \begin{align}
 \label{cont1}
 \frac{\sin\frac{\theta}{2}}{\sinh\frac{\gamma_{\theta}+\gamma_0}{2}}\rightarrow
 \frac{\sinh2\kx}{\sinh2\ky}\frac{\sinh\frac{\gamma_{\theta}+\gamma_0}{2}}{\sin\frac{\theta}{2}},
 \quad e^{\nu_{\theta}/2}\rightarrow e^{\nu_{\theta}/2}\sqrt{\frac{\sinh2\kx}{\sinh2\ky}}
 \frac{\sinh\frac{\gamma_{\theta}+\gamma_0}{2}}{\sin\frac{\theta}{2}},
 \end{align}
 for $\theta\in(0,2\pi)$ and
  \begin{align}
  \label{cont2}
  \frac{e^{-\nu_0/2}}{\sqrt{N\sinh\gamma_0}}\rightarrow &\, \sqrt{\frac{\sinh2\kx}{\sinh2\ky}}\left(\frac{e^{-\nu_0/2}}{\sqrt{N\sinh\gamma_0}}\right)^{-1},
  \\
  \label{cont3} \xi\xi_T\rightarrow &\,\xi\xi_T \sqrt{\frac{\sinh2\ky}{\sinh2\kx}}\frac{e^{-\nu_0}}{N\sinh\gamma_0}.
  \end{align}
 Quasiparticle interpretation of eigenvectors also changes. If the particle with $\theta=0$ was
 initially present or absent in a $p$-eigenstate, then it is annihilated (resp. created) after analytic
 continuation. This means e.g. that the number of particles in $p$-eigenstates of the periodic Ising transfer matrix becomes odd instead of even. Relations (\ref{cont1})--(\ref{cont3}) then show that
 spin form factors are given by the same formula
 (\ref{finalff}), with odd $m+n$ and appropriately adjusted first numerical factor.

 It is also possible to derive this result along the above lines. However, for $\kx^*>\ky$
 the definition of the creation-annihilation operators (\ref{caops}) should be modified
 and the corresponding matrix $D$ is degenerate, which leads to additional subtleties.
 \end{remark}

 \section{Discussion}
 In this paper, exact expressions for finite-lattice form factors of the spin operator
 in the two-dimensional Ising model were obtained. Starting point of our derivation was the idea of expressing induced linear transformations of fermions in a particular basis, which was
 put forward by Hystad and Palmer in \cite{Hystad,Palmer_Hystad}. New crucial ingredient is the use of elliptic determinants and theta functional interpolation.
 The presented approach seems to be quite natural and straightforward.
 It is likely to be applicable to other free-fermion lattice models, such as those considered
 in \cite{Iorgov5,OL,smj}.

 A more ambitious challenge is to complete the program of form factor derivation for
$\mathbb{Z}_N$-symmetric superintegrable chiral Potts quantum chain, which presumably possesses a hidden fermion structure. Two parts of the hamiltonian of this model generate Onsager algebra and the space of states decomposes into a set of invariant irreducible subspaces (Onsager sectors) with Ising-like
hamiltonian spectrum in each of them.

Using Baxter's idea of  extending Onsager algebra by the spin operator \cite{Baxter2},
its form factors between the hamiltonian eigenstates were found in
\cite{Iorgov4} up to unknown scalar factors for each pair of Onsager sectors.
In the case $N=2$ (quantum Ising chain in a transverse field),
 Onsager algebra can be embedded into a fermion algebra, so that all
 irreducible representations of the former are combined into one irreducible representation of
 the latter. This allows to fix all unknown scalar factors and obtain
 form factors of the quantum Ising chain \cite{Iorgov3}. It is therefore natural to look for
  a fermion-like algebra extending Onsager algebra for general $N$.

    \begin{acknowledgements}
  The authors are grateful to G. von Gehlen and S. Pakuliak
  for stimulating discussions and especially to S. Spiridonov for drawing their attention
  to Example~3 on p.~451 of \cite{WW}.
  We also thank Max Planck Institute for Mathematics (Bonn),
  where a part of this research was done, for hospitality and excellent working conditions.
  This work was supported by the
  French-Ukrainian program Dnipro M17-2009, the joint PICS project  of CNRS and NASU,
  IRSES project ``Random and Integrable Models in Mathematical Physics'', the Program of Fundamental Research of the Physics and Astronomy
 Division of the NASU and Ukrainian FRSF grants $\Phi$28.2/083 and $\Phi$29.1/028.
  \end{acknowledgements}

 \end{document}